\documentclass[prd,amsfonts,onecolumn,superscriptaddress,aps,nofootinbib,11pt]{revtex4-1}

\usepackage{amsmath,amssymb,mathrsfs}
\usepackage{graphicx}
\usepackage{enumerate}
\usepackage[colorlinks=true]{hyperref}
\pdfoutput=1

\usepackage[]{units}
\usepackage{ulem} 
\providecommand{\eq}[1]{\begin{equation} #1 \end{equation}}
\providecommand{\eqali}[1]{\begin{equation}\begin{aligned} #1
    \end{aligned}\end{equation}}
\providecommand{\lag}{\mathcal{L}}

\providecommand{\ums}[2][1]{\ml{\tfrac{#1}{#2}}}
\providecommand{\bl}{\bar{\lambda}}
\providecommand{\ZZ}{\mathbb{Z}}
\providecommand{\ml}[1]{\mbox{\large $#1$}}
\providecommand{\mss}[1]{\mbox{\scriptsize $#1$}}
\providecommand{\aver}[1]{\langle #1 \rangle}
\providecommand{\tp}{{\mss{\mathsf{T}}}}
\providecommand{\mtrx}[1]{\begin{pmatrix} #1 \end{pmatrix}}

\providecommand{\pq}{\mathrm{PQ}}

\providecommand{\to}{\rightarrow}
\newcommand{\misset}{\not\!\! E_T}

\begin{document}
\title{Collider and Dark Matter Searches in the Inert 
Doublet Model from Peccei-Quinn Symmetry
}

\author{Alexandre Alves}%
\email{aalves@unifesp.br}
\affiliation{
Departamento de Ci\^encias Exatas e da Terra, Universidade Federal de S\~ao Paulo, 
Diadema-SP, 09972-270, Brasil
}

\author{Daniel A. Camargo}%
\email{dacamargov@gmail.com}
\affiliation{Universidade Federal do ABC, Centro de Ci\^encias Naturais e Humanas, 09210-580, Santo Andr\'e-SP, Brasil}

\author{Alex G. Dias}%
\email{alex.dias@ufabc.edu.br}
\affiliation{Universidade Federal do ABC, Centro de Ci\^encias Naturais e Humanas, 09210-580, Santo Andr\'e-SP, Brasil}

\author{Robinson Longas}%
\email{robinson.longas@udea.edu.co}
\affiliation{Instituto de F\'isica, Universidad de Antioquia, Calle 70 No. 52-21, Medell\'in, Colombia}

\author{Celso C. Nishi}%
\email{celso.nishi@ufabc.edu.br}
\affiliation{Universidade Federal do ABC, Centro de Matem\'atica, Computa\c{c}\~ao 
e Cogni\c{c}\~ao Naturais,  09210-580, Santo Andr\'e-SP, Brasil}

\author{Farinaldo S. Queiroz}%
\email{farinaldo.queiroz-hd.mpg.de} 
\affiliation{Max-Planck-Institut fur Kernphysik, Saupfercheckweg 1, 69117 
Heidelberg, Germany}

\date{\today}

\begin{abstract}
Weakly Interacting Massive Particles (WIMPs) and axions are arguably the most 
compelling dark matter candidates in the literature.  Could they coexist as dark 
matter particles? More importantly, can they be incorporated in a well motivated 
framework in agreement with experimental data? In this work, we show that this two 
component dark matter can be realized in the Inert Doublet Model in an elegant 
and natural manner by virtue of the spontaneous breaking of a Peccei-Quinn
$U(1)_{PQ}$ symmetry into a residual $\ZZ_2$ symmetry. The WIMP stability is 
guaranteed by the $\ZZ_2$ symmetry and a new dark matter component, the axion, 
arises. There are two interesting outcomes: (i) vector-like quarks needed to 
implement the Peccei-Quinn symmetry in the model act as a portal between the dark 
sector and the SM fields with a supersymmetry-type phenomenology at colliders; (ii) 
two-component Inert Doublet Model re-opens the phenomenologically interesting 
100-500 GeV mass region. We show that the model can plausibly have two component 
dark matter and at the same time avoid low and high energy physics constraints such 
as monojet and dijet plus missing energy, as well as indirect and direct dark 
matter detection bounds.
\end{abstract}
\maketitle
\section{Introduction}

It is quite possible that the dark matter (DM), amounting to approximately 27\% 
of the total energy density of the Universe,
may be constituted by more than one particle.
One of the most popular candidate for DM is the generic Weakly Interacting Massive 
Particle (WIMP) that suggests the connection between DM physics and the weak 
scale.
The stability of the WIMP is usually assumed to be due to the presence of a 
discrete global symmetry, such as a $\ZZ_2$ symmetry, which prevents its decay.  
Another candidate is the axion~\cite{Weinberg:1977ma, Wilczek:1977pj}, which is the  
pseudo Nambu-Goldstone of the breakdown of the $U(1)_{PQ}$ Peccei-Quinn (PQ) 
symmetry proposed to solve the strong CP problem~\cite{Peccei:1977hh}
(see Refs.~\cite{Kim:2008hd,Jaeckel:2010ni,Ringwald:2012hr} for a review). 
Under the assumption that the $U(1)_{PQ}$ symmetry is broken at an energy scale 
much higher than the electroweak scale, the axion can be an ultralight particle 
with faint interactions with all other 
particles~\cite{Kim:1979if,Shifman:1979if,Dine:1981rt,Zhitnitsky:1980tq}, and 
allowed to have a lifetime larger than the age of the Universe. The axion 
contribution to the total DM energy density in the Universe also depends on the 
energy scale in which the $U(1)_{PQ}$ symmetry is broken~\cite{Sikivie:2006ni}. 
Thus, the scenario in which both WIMP and axion make up the DM of the Universe is a 
natural and compelling framework.
With that in mind we add a new and well motivated ingredient, the axion, on
one of the simplest extensions of the Standard Model with a WIMP: the Inert Doublet 
Model (IDM), which contains an additional $SU(2)_L$ Higgs doublet with the lightest 
component stabilized by an {\it ad hoc}  $\ZZ_2$ 
symmetry~\cite{LopezHonorez:2006gr,Honorez:2010re,LopezHonorez:2010tb}.

In other words, we propose the axion as the DM companion to the IDM 
component $H^0$. 
To this end we have developed a model based on the observation made 
in~\cite{Dasgupta:2013cwa}, where a $U(1)_{PQ}$ symmetry broken spontaneously into a 
$\ZZ_2$ symmetry was advocated to stabilize the
WIMP\,\footnote{Other contexts where the WIMP is stabilized by an 
accidental symmetry that remains from the breaking of a more fundamental 
symmetry at a higher energy scale are given
in~\cite{Kadastik:2009dj,Kadastik:2009cu,Frigerio:2009wf,Nagata:2015dma, 
Arbelaez:2015ila,Boucenna:2015sdg,Heeck:2015qra}
}.
We tacitly assume that the $U(1)_{PQ}$ symmetry is protected against gravitational 
effects -- which generate Planck-scale-suppressed symmetry breaking operators --
by some sort of discrete symmetry (as in e.g. \cite{Dias:2014osa,Ringwald:2015dsf})
to avoid destabilization of the solution to the strong CP problem, and also of the 
WIMP \cite{Mambrini:2015sia}.
The use of this global symmetry to stabilize the WIMPs is safe from gravitational 
effects which might violate the $U(1)_{PQ}$ \cite{Mambrini:2015sia}, since only 
Planck suppressed operators of dimension six are present.
To complete this two component DM system, at least a scalar singlet field 
hosting the axion $a$ and a vector-like quark $D$ are necessary 
in addition to the inert Higgs doublet whose lightest neutral component is the heavy 
DM~\cite{Deshpande:1977rw,Ma:2006km,Barbieri:2006dq}.
The vector-like quark allows a simple 
implementation of the $U(1)_{PQ}$ symmetry, as in the KSVZ axion 
model~\cite{Kim:1979if,Shifman:1979if}, and  acts as a portal connecting the SM and 
the dark sector. As a consequence of the residual  $\ZZ_2$ symmetry, the heavy 
vector quarks decay only to new heavy scalars and SM quarks, mimicking the 
phenomenology of $R$-parity conserving supersymmetry (SUSY) at colliders, including 
the classic SUSY signal of jets 
plus large missing energy.

As there is currently many experimental constraints on supersymmetry from the LHC 
searches, we performed, prior to the study of the multi-component DM 
scenario of our model, an investigation of the limits from the searches of jets plus 
missing energy and monojets at the LHC. After that, we  focused on the main goal of 
the paper, which is the study of our axion-WIMP DM scenario, pointing out 
the differences in relation to the typical IDM. The main finding is that, in 
contrast with the one-component DM in the IDM, the phenomenologically 
important mass interval $100\,{\rm GeV} \leq M_{H^0}\leq 500\,{\rm GeV} $ is 
re-opened, with the axion filling the role of the remaining DM.

This paper is organized as follows. In section~\ref{sec:model} we present the model 
and show the particle spectrum. In section~\ref{sec:Pheno} we show that the model is 
consistent with the actual constraints from searches of events having signatures of 
jets + missing energy/monojets at the LHC. Then, in section~\ref{sec:DM} we discuss 
the implications of having a axion-WIMP mixed DM scenario and study the 
coannihilations with the exotic quarks due to the new vector-like portal. In 
addition, we discuss the constraints imposed by direct and indirect detection 
searches for WIMPs. The conclusions are presented in 
section~\ref{sec:concluisions}.

\section{The model}
\label{sec:model}

The model consists on a KSVZ type axion model~\cite{Kim:1979if,Shifman:1979if} with an inert doublet $H_D$, whose the 
lightest neutral component is stabilized by a residual $\ZZ_2^D$ symmetry that remains unbroken from the 
original PQ symmetry. 
Therefore, we will have two candidates for DM: the ultralight axion and the 
WIMP-like lightest component of $H_D$.

The simplest way to implement the breaking $U(1)_\pq\to \ZZ_2^D$ is to break the 
PQ symmetry by a vev of a singlet scalar $S$ of $\pq(S)=2$ while all other fields 
carry integer PQ charges. The fields carrying even or zero PQ charge will be even 
under the remaining $\ZZ_2^D$ whereas those carrying odd PQ charge will be odd under 
$\ZZ_2^D$, and thus belong to the dark sector.
The conservation of $\ZZ_2^D$ requires that scalars with odd PQ charge be 
inert.
As usual, the responsible for PQ symmetry breaking will host the axion in its phase 
as 
\eq{
S=\frac{1}{\sqrt{2}}(f_a + \rho(x))e^{i a(x)/f_a}\,,
}
where $a(x)$ is the axion field, and $f_a$ the axion decay constant that corresponds to the vev of $S$ in our case (a KSVZ type axion model~\cite{Kim:1979if,Shifman:1979if}). 
Nonperturbative QCD effects lead to a potential, which generates a mass to the 
axion as
\begin{equation}
m_a\approx 6 \,\unit{meV}\times (10^9\,\unit{GeV}/f_a)\,.
\end{equation}

In this framework the axion couplings with matter and gauge bosons are suppressed by 
$f_a$
which, being much higher than the electroweak scale, makes the axion an ultralight 
particle with feeble couplings to all other particles.
In fact, $f_a$ is constrained from astrophysical objects which would have their 
dynamics affected if axions interact too much with photons.
For example, supernova SN1987A data constrains $f_a$ to be greater than $10^9$ 
GeV~\cite{Raffelt:1987yt,Essig:2013lka}. Still, an upper limit on the  decay 
constant is obtained from the requirement that the axion relic density should not 
exceed the DM density, which gives $f_a\leq 10^{12}$ 
GeV~\cite{Bae:2008ue,Visinelli:2009zm,Wantz:2009it,Hertzberg:2008wr}. 

In addition to the SM fermions we assume that there is at least one heavy quark field
$D\sim ({\bf 1},\,-1/3)$, where the numbers inside the parenthesis 
represent the transformation properties under the electroweak gauge group factors 
$SU(2)_L$ and $U(1)_Y$; the case of charge $2/3$ exotic quark can be treated 
analogously. 
Such a quark field is formed by left- and right-handed fields 
$D_{L,R}$, having the following interaction with $S$ 
\eq{
\lag\supset yS^*\overline{D}_L D_R+h.c.\,,
\label{lad}
}
so that $\pq(D_{L})=-1$ and $\pq(D_{R})=1$. This results in a nonzero value for the 
anomaly coefficient, $c_{ag}=\pq(D_L)-\pq(D_R)=-2$, allowing the axion to have a 
coupling with the gluon field strength as required to solve the strong CP problem 
through the Peccei-Quinn mechanism.

With the vev of $S$ a mass $M_D= y\,f_a/\sqrt{2}$ for the $D$ quark is generated 
through the interaction in Eq. (\ref{lad}). We tune the Yukawa coupling $y\leq 
10^{-6}$ in Eq. (\ref{lad}), as $f_a\geq 10^9$ GeV, so that $M_D$  lies at the 
TeV scale. In the appendix \ref{ap:uv-qft} it is shown how to ameliorate such a 
tuning by extending the model.

Besides the fields necessary to solve the strong CP problem, we augment the SM with 
an inert Higgs doublet $H_D\sim ({\bf 1},\,1/2)$, with $\pq(H_D)=-1$, 
in addition to the usual Higgs doublet $H\sim ({\bf 1},\,1/2)$.
In the limit of PQ symmetry conservation, the Higgs potential is effectively\,%
\footnote{We consider that the effective parameters already includes the effects of 
integrating out the heavy fields at the PQ scale.}
\begin{align}
  \label{eq:potential}
  V & =  \mu_1^2 H^{\dagger}H + \mu_2^2 H_D^{\dagger}H_D
  + \frac{\lambda_1}{2} ( H^{\dagger}H)^2
  + \frac{\lambda_2}{2}  (H_D^{\dagger}H_D)^2
  +  \lambda_3 ( H^{\dagger}H )( H_D^{\dagger}H_D )
  +  \lambda_4 |H^{\dagger}H_D|^2\,.
\end{align}
Exact PQ symmetry at the electroweak scale would imply degenerate CP odd and CP 
even scalars of the inert doublet, a feature that is problematic if the inert 
doublet accounts for all or most of the DM: 
direct detection searches for inelastic DM requires a mass splitting larger than
100\,\unit{keV}\,\cite{Hambye:2009pw, Barbieri:2006dq}. As PQ symmetry is broken at the scale $f_a$ we expect the additional PQ-violating but $\ZZ_2^D$ conserving term to be generated:
\eq{
\label{lambda5}
\delta V=\ums{2}\lambda_5(H^\dag H_D)^2+h.c.
}
The mass splitting is thus controlled by $\lambda_5$, which can be taken real.
A simple completion  that generates the term \eqref{lambda5} is shown in appendix 
\ref{ap:uv}.
The fields beyond the SM, along with their quantum numbers, are collected in Table \ref{tab:content}.
\begin{table}[h!]
\begin{center}
\begin{tabular}{|c|c|c|c|c|}
\hline\rule[0cm]{0cm}{.9em}
  & $D_L$ & $D_R$ & $H_D$ & $S$ \\
\hline
\hline
$SU(3)_C$ & $\bf{3}$ & $\bf{3}$ & $\bf{1}$ & $\bf{1}$  \\
\hline
 $SU(2)_L$ & $\bf{1}$ & $\bf{1}$ & $\bf{2}$ & $\bf{1}$  \\
 \hline
 $U(1)_{PQ}$ & $-1$  & $1$ & $-1$  & $2$  \\ \hline
 $\ZZ_2^D$ & $-$  & $-$ & $-$  & $+$  \\ 
\hline
\end{tabular}
\end{center}
\caption{Quantum numbers of the fields beyond the SM.
}
\label{tab:content}
\end{table}
The interaction between the dark sector and the SM will be given essentially from 
the Yukawa term (apart from the interaction term involving the standard Higgs boson 
and the inert Higgs doublet in the potential), acting as an inert doublet portal,
\begin{align}
\label{inert.portal}
\lag\supset y_D\overline{q_{L}} H_D D_R + {\rm h.c}\,. 
\end{align}
where $q_{L}\sim ({\bf 2},\,1/6)$, corresponds to the three families of SM doublets 
of quarks and $y_D$ is the Yukawa coupling. 
We will effectively consider that there is only one heavy vector-like quark $D$ 
accessible to the LHC and relevant to DM coannihilations. 
The possible constraints coming from these processes and also from the DM direct 
detection will be one of our goals.
Moreover, we will choose this TeV scale heavy quark to couple only to 
one family of SM quarks.
This choice will suppress new flavor violating effects such as on 
$D^0-\bar{D}^0$. In particular, the case in which the exotic quark couples 
only to the first quark family follows by imposing minimal flavor violation: for 
three families of heavy quarks $D_{iL,R}$ with $D_{iL}\sim d_{iR}$ 
($D_{iL}\sim u_{iR}$) and $D_{iR}\sim q_{iL}$ the spectrum for $D_i$ can be chosen 
hierarchical as the SM down (up) quarks and with same order and approximately 
diagonal Yukawa couplings (as studied in, 
e.g., Refs.\,\cite{Arnold:2010vs,Grossman:2007bd}, with the difference that in 
our case the light-heavy quark mixing is absent due to $\ZZ_2^D$).
We obtain only one heavy quark interacting predominantly to the first 
family after integrating out the much heavier fields.%
\,\footnote{In this case, the axion-photon coupling should change appropriately.}
The other cases are considered for phenomenological comparison.

The spectrum at the electroweak scale which we consider is an inert doublet model~\cite{Barbieri:2006dq,LopezHonorez:2006gr} augmented by an axion and a vector-like quark $D$ interacting with the particles of the SM  
through Eq.~\eqref{inert.portal}. The dark sector, odd by $\ZZ_2^D$, consists of the fields of the inert doublet 
$H_D$ and the vector quark $D$. We choose the lightest component of $H_D$ to 
be lighter than $D$ and then be part of the DM content along with the axion.
It has to be noted that several models at the PQ scale can lead to this spectrum at low 
energies. A simple complete model that leads to this spectrum is shown in appendix 
\ref{ap:uv}; it coincides with model I of Ref.~\cite{Dasgupta:2013cwa} but with a 
different spectrum at low energies.

The electroweak symmetry breaking is still performed by 
$\aver{H}=v/\sqrt{2}(0,1)^\tp$, where $v=246\,\unit{GeV}$, with the resulting CP even state from the doublet $H$, identified as the standard Higgs boson, denoted by $h$, with mass $m_h= 125\,\unit{GeV}$. The components of the inert doublet 
\eq{
H_D=(H^+,\frac{H^0+iA^0}{\sqrt{2}})^\tp\,,
}
give rise to four physical states: a charged state $H^+$ and its charge conjugate, a neutral 
and CP odd $A^0$, and a neutral and CP even $H^0$.
Note that $H^0$ does not develop a vacuum expectation value  in 
order to leave the remnant $\ZZ_2^D$ symmetry unbroken. Thus, the scalar potential 
gives rise to the quartic interaction $\ums{2}\lambda_Lh^2X^2$ where $X$  is the 
lightest between $H^0$ or $A^0$ and $\lambda_L\equiv 
\ums{2}(\lambda_3+\lambda_4-|\lambda_5|)$, which quantifies the strength of the 
Higgs portal.

After the spontaneous symmetry breaking the scalars acquire the masses
\eqali{
\label{eq:scalar_neutralsmasss}
M^2_{H^\pm}&=\mu_2^2+\ums{2}\lambda_3v^2\,,\cr
M^2_{H^0}&=\mu_2^2+\ums{2}\lambda_{345}v^2\,,\cr
M^2_{A^0}&=\mu_2^2+\ums{2}\bl_{345}v^2\,,
}
where $\lambda_{345}\equiv\lambda_3+\lambda_4+\lambda_5$ and $\bl_{345} 
\equiv\lambda_3+\lambda_4-\lambda_5$.
We can see that the scalar-pseudoscalar mass splitting is indeed controlled by 
$\lambda_5$:
\begin{align}
M_{H^0}^2-M_{A^0}^2 =\lambda_5 v^2\,.
\end{align}

In summary, the model has eight free parameters namely,
\eq{
\label{free}
\{M_{H^\pm},M_{H^0},M_{A^0},M_D ,y_D ,\lambda_2,\lambda_L,f_a\}\,,
}where the first four elements in this set are the masses of the  particles which are odd under $\ZZ_2^D$, with $\lambda_5<0$ guaranteeing $H^0$ to be the lightest scalar of the dark sector besides the 
axion. The case in which $A^0$ is the lightest CP odd scalar is directly obtained   
replacing $\lambda_5\to -\lambda_5$.
As we describe in what follows, these parameters will be subjected to a multitude 
of constraints from the electroweak nature of the model which will reduce the 
viable parameter space considerably.
These include theoretical constraints as well as various phenomenological ones.

\smallskip
{\bf Vacuum Stability and Perturbativity}

Considerations such as vacuum stability and perturbativity restrict the range of 
parameters in~\eqref{free}.
For the potential to be bounded from below, we 
need~\cite{Deshpande:1977rw,Ivanov:2006yq}
\eq{
\lambda_1\ge 0, ~~
\lambda_2\ge 0, ~~
\lambda_3+\sqrt{\lambda_1\lambda_2}> 0, ~~
2\lambda_L+\sqrt{\lambda_1\lambda_2}> 0\,.
}
To ensure the inert minimum ($\aver{H}=v/\sqrt{2}(0,1)^\tp$, $\aver{H_D}=(0,0)^\tp)$ to be the global minimum we 
require~\cite{Swiezewska:2012ej}
\eq{
(\text{scalar masses})^2\ge 0\,,~~
\frac{\mu^2_1}{\sqrt{\lambda_1}}<
\frac{\mu^2_2}{\sqrt{\lambda_2}}\,.
}
In special, the positivity of the usual Higgs mass squared requires $\mu^2_1<0$.
When one-loop effects are considered \cite{Ilnicka:2015jba}, this condition may not be 
strict~\cite{Ferreira:2015pfi}.
We also require perturbativity of the scalar quartic couplings, assuming~\cite{Ilnicka:2015jba}
\eq{
|\text{quartic self-couplings}|<4\pi\,,~~
|\text{$X^\dag Xhh$ coupling}|<4\pi. 
\label{lpert}
}
Applied to the $(H^0)^4$ coupling, the first requirement\,%
\footnote{Within the IDM, the second requirement in~\eqref{lpert} leads to an upper bound for scalar masses of tenths of TeV if the correct relic abundance for $H^0$ is 
required~\cite{Hambye:2009pw}.
}
in~\eqref{lpert} 
translates into $\lambda_2\le \frac{4}{3}\pi\approx 4.19$~\cite{Ilnicka:2015jba}.
A related constraint would be the unitarity in the scalar-scalar scattering 
matrix~\cite{Ginzburg:2005dt}.
We do not impose the latter explicitly and argue that perturbativity already cuts 
off most of the non-unitary cases.

\smallskip
{\bf Electroweak Bound}

The first basic constraint comes from the electroweak nature of $H_D$ and requires 
that the SM gauge bosons cannot decay into the dark scalars, i.e.,
\eq{
M_{H^0}+M_{A^0}>m_Z,\quad\,
M_{H^0}+M_{H^\pm},\,
M_{A^0}+M_{H^\pm}>m_W\,.
}

\smallskip
{\bf LEP Limit}

Susy searches at LEP~\cite{Lundstrom:2008ai} further exclude
$M_{H^0}<80\,\unit{GeV}$ and $M_{A^0}<100\,\unit{GeV}$, for 
$M_{A^0}-M_{H^0}>8\,\unit{GeV}$, for the neutral scalars and 
$M_{H^\pm}<70\,\unit{GeV}$ for the charged one.

\smallskip
{\bf LHC - Higgs Invisible Width}

Additionally, when $M_{H^0}<m_h/2$, invisible Higgs decays put strong constraints 
on the Higgs portal coupling,
\eq{
|\lambda_L|\lesssim 0.012 ~(0.007)\,,
}
for $M_{H^0}=60\,\unit{GeV}$ ($M_{H^0}=10\,\unit{GeV}$) when only $h\to H^0H^0$ is 
open \cite{Belanger:2013xza}.
 Thus we choose hereafter 
 \eqali{
 \label{param:range}
 M_{H^\pm},M_{A^0} > 100\,\unit{GeV},\;
 M_{H^0}>60\,\unit{GeV}.
 }
 
\smallskip
{\bf LHC - Dilepton + Missing Energy Data}

Using dilepton plus missing energy data from the LHC, bounds have been placed in the 
IDM for $M_{H^0} < M_W$ (the $W$ boson mass), based on production channels such as 
$q\bar{q} \rightarrow 
Z \rightarrow A^0 H^0 \rightarrow Z^{\star} H^0 H^0 \rightarrow l^+l^- H^0 H^0$. In 
\cite{Belanger:2015kga} the authors were able to rule out $H^0$ masses below 
$35$~GeV at 95\% C.L. with Run I data. Thus far, the Higgs resonance region, where 
the relic density, direct, and indirect detection bounds are satisfied is left 
untouched. Anyway, this mass region lies outside our scenario in 
\eqref{param:range}. (See \cite{Dolle:2009ft} for an old study of dilepton data in 
the IDM).

We have reviewed the key aspects of the model as well as existing constraints for 
the IDM. Hereunder we discuss collider constraints based on monojet and dijet plus 
missing energy data from LHC at $7-8$~TeV.

\section{Collider constraints}
\label{sec:Pheno}

By virtue of the $\ZZ_2^D$ symmetry, the vector-like quarks can only decay into a quark and a new heavy scalar, including 
the DM $H^0$. Pair production of these new heavy quarks gives rise to 
SUSY-like signatures at colliders as jets plus missing energy, while associated 
production of a heavy quark and $H^0$ leads to monojets. 
Therefore, constraints from collider searches for supersymmetry and DM have to be 
taken into account prior to a dedicated study of our DM candidate. Let us discuss 
how we checked these collider bounds.

\subsection{Bounds from SUSY and DM searches in jets plus missing energy and monojets}

\begin{figure}[t]
 \includegraphics[scale=0.61]{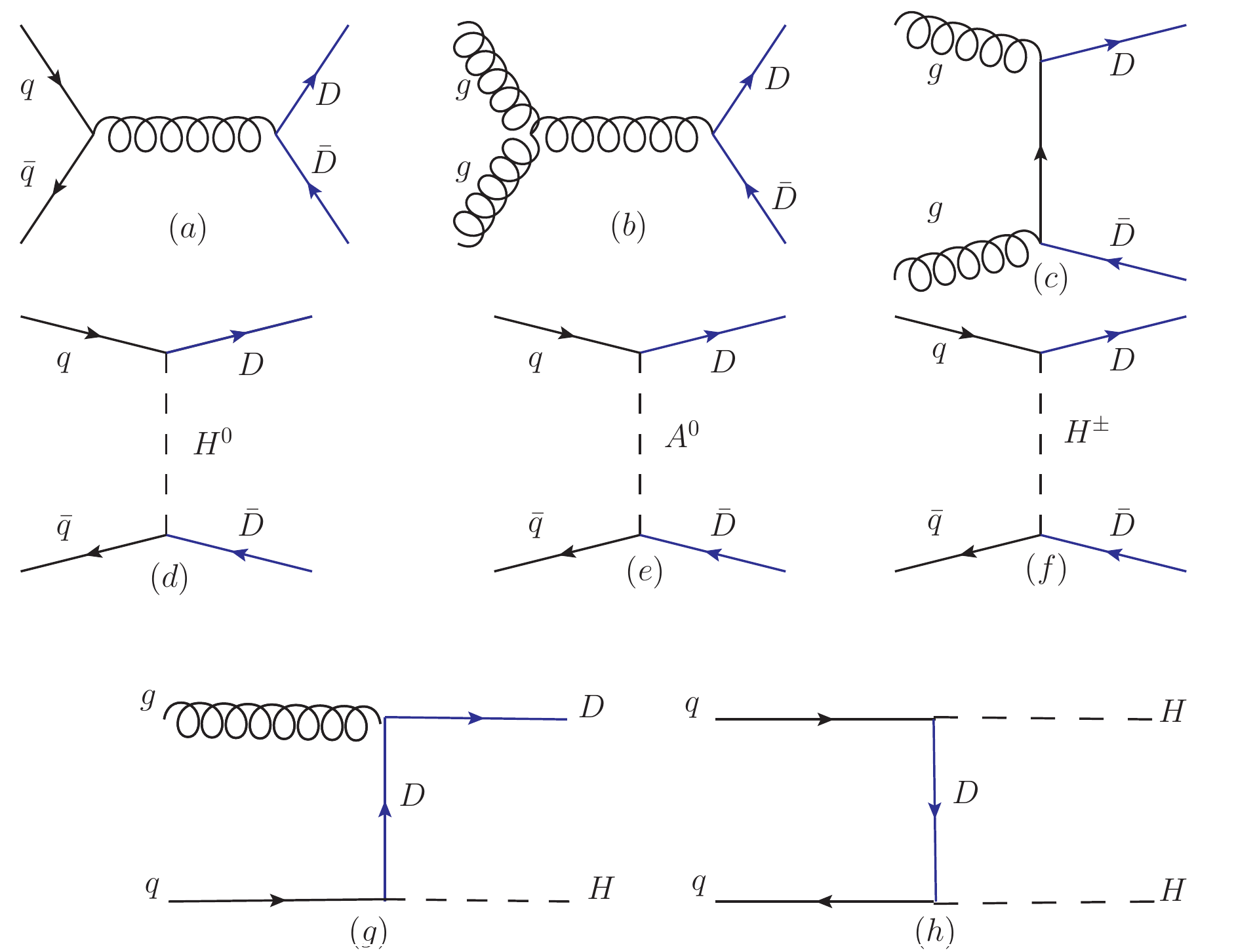}
\caption{Feynman diagrams for production of $D\overline{D}$ pairs
in proton-proton collisions are shown in diagrams (a)--(f). Additional diagrams obtained from crossing or charge conjugation of 
the initial and final state are not shown. In diagram (g) we display one contribution to $D+H^0$ associated production. 
Diagram (h) represents a subleading contribution to monojet signatures when a QCD jet from a strongly interacting line is radiated. }
\label{fig:production}
\end{figure}

As aforementioned, due to the $\ZZ_2^D$ symmetry, the vector-like quark $D$ can always be pair produced ($D\overline{D},DD,\overline{D}\overline{D}$) via quark or gluon fusion, or in association with a new scalar ($H^{0},A^{0},H^{\pm}$) as shown in Fig.~\ref{fig:production}. In particular, in Fig.~\ref{fig:production} 
we display representative contributions for pair production, diagrams (a)--(f), and single production in association with $H^0$, diagram (g). 

Singlet vector-quarks $D$ can interact with the down, strange and bottom quarks via Yukawa couplings to the scalars of the model. These Yukawa couplings might be constrained by flavor physics and searches for new physics in colliders.
For example, low energy physics impose constraints on the Yukawa couplings for the
case where $D$ couples with more than one family of SM quarks. We thus adopt safe 
benchmarks to render the model free from constraints on quark flavor violation 
allowing $D$ to interact just with one family of SM quarks at a time through the 
Yukawa coupling $y_{D}$.

For the pair production of $D$, both QCD and Yukawa interactions with the scalars $H^\pm,\; A^0,\; H^0$ contribute to the cross section.
The $t$-channel diagrams with neutral scalars allow for $DD$ and 
$\overline{D}\overline{D}$ production alongside $D\overline{D}$; see diagrams (d) 
and (e) in Fig.~\ref{fig:production}. 
A similar situation occurs in squark pair production where $t$-channel gluinos 
contribute to same-sign squarks production. 
Also, as in the case of squarks, the $t$-channel contributions impact significantly the production cross section of jets and missing energy. 

\begin{figure}[!h]
\includegraphics[scale=0.51]{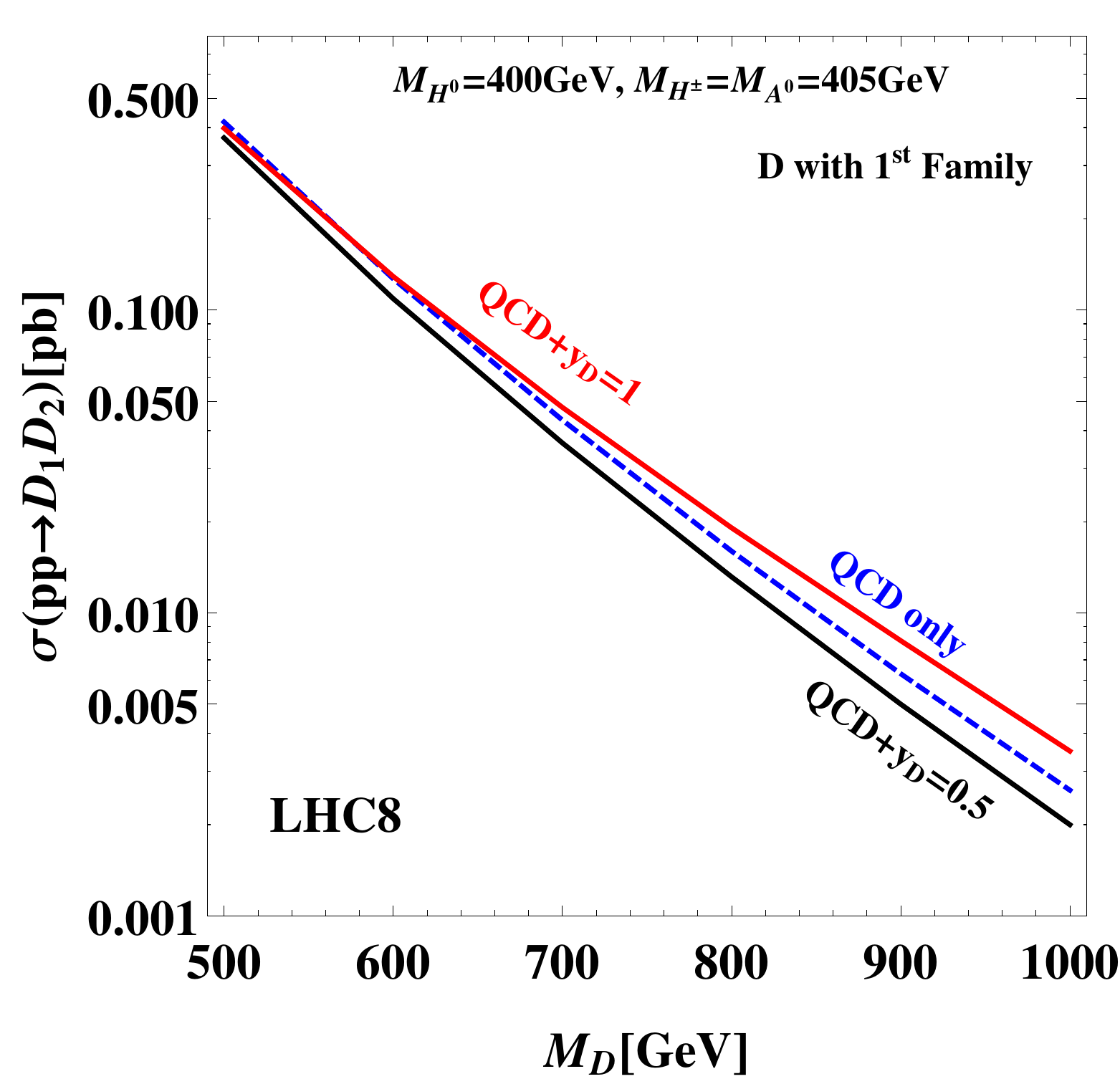}
\includegraphics[scale=0.51]{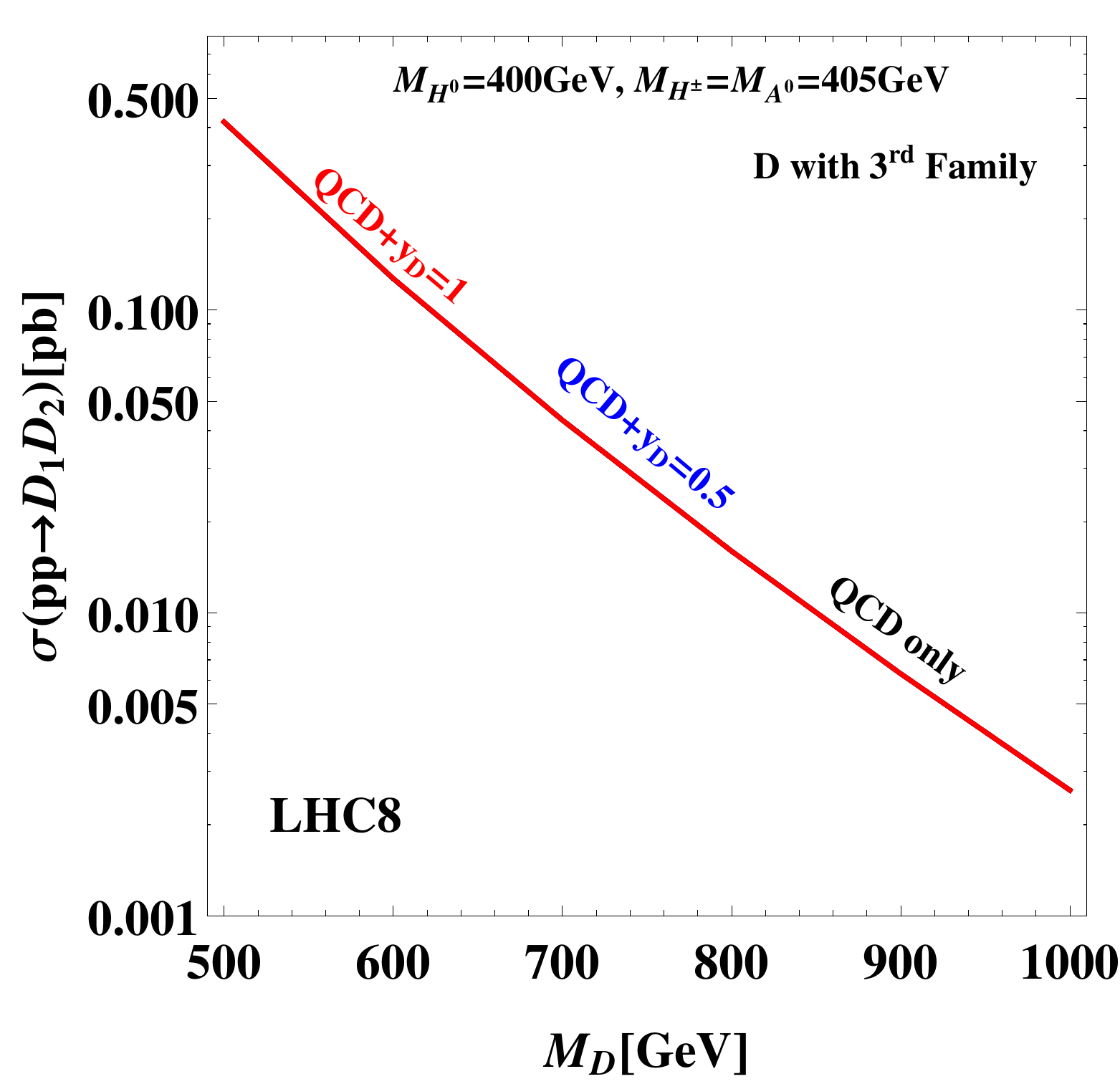}
\caption{Pair production cross section of $D$ quark at the $\sqrt{s}=8$ TeV LHC for $D$ coupling with 
the first (left panel), and third (right panel) SM-generations. The result for the second family is identical to the third. The sum of the cross sections for the production of opposite-charge ($D\overline{D}$) 
and same-charge quarks ($DD+\overline{D}\overline{D}$) as a function of the vector quark mass are displayed in solid lines.
The red (black) solid line assumes $y_D=1$ (0.5). The blue dashed line is the QCD contribution for $D\overline{D}$ production. 
The scalars masses are fixed as $M_{H^{0}}=400\hspace{0.2cm}\hbox{GeV}$ and $M_{A^{0}}=M_{H^{\pm}}=405\hspace{0.2cm}\hbox{GeV}$. 
}
\label{fig:xsex}
\end{figure}

It is shown in Fig.~\ref{fig:xsex} the pair production cross section $\sigma(pp\to D_1D_2)$ for the 8 TeV LHC for couplings with the first (down) and third (bottom) quark families, where $D_{1(2)}$ represents both a heavy quark and a heavy antiquark. 
The solid red (black) line represents the total cross section with all contributions from QCD and Yukawa couplings setting 
$y_D=1$ (0.5), $M_{H^{0}}=400$ GeV and $M_{A^{0}}=M_{H^{\pm}}=405$ GeV. The pure QCD 
contribution is shown as a dashed blue line. Interestingly, the interference between 
the QCD and the $t$-channel Yukawa contributions is destructive, contrary to the 
SUSY case.
The interference is visible only for the case of couplings with the first family, 
as shown in Fig.~\ref{fig:xsex} where we can see at the left (right) panel the 
production cross section of vector-quark pairs with $d$($b$)--$D$--$H^0$ coupling 
only.
This is, of course, due to the parton content of the proton; the non-QCD 
$t$-channel diagrams connect only the initial state quarks participating in the 
Yukawa coupling to the vector quark $D$, thus, scenarios with exclusive couplings to 
the second and third families are suppressed and the Yukawa amplitudes contribute 
too little. For moderate Yukawa couplings $y_D\lesssim 0.5$, the destructive 
interference decreases the total cross section and only at larger Yukawa coupling 
regimes, where $y_D\ge 1$, the production rate can become larger than the pure QCD 
contribution.

The single and pair production of the quark $D$ lead to monojet and two jets plus missing energy signatures at the LHC, respectively. 
Monojets also receive contributions from diagram (h) of Fig.~\ref{fig:production} when a QCD jet is radiated from a strongly interacting 
particle. Monojets are striking signatures expected in the case that DM is produced 
in proton-proton collisions while two or 
more jets plus missing energy is the classical signature for production and decay of squarks and gluinos. Upper limits for production 
cross section times branching ratios for processes with hard jets and missing energy have been placed by the ATLAS and CMS 
Collaborations in the 7 and 8 TeV run of the LHC, and incorporated to the database 
of packages aimed to check for collider 
limits as \texttt{SmodelS}~\cite{Kraml:2013mwa} and \texttt{CheckMate}~\cite{Drees:2013wra}. 

As the quark $D$ has a decay channel into a jet and $H^0$, both the constraints from 
squark searches and DM searches apply in 
our case. In order to check these bounds we simulated the collision processes
\begin{align}
pp & \to D\overline{D}(DD)(\overline{D}\overline{D})\to jj+\misset\label{jjMET}\\
pp & \to D(\overline{D})H^0\to j+\misset\label{jMET1} \\
pp & \to H^0H^0+j\to j+\misset\label{jMET2} 
\end{align}
up to one extra jet to approximate higher order QCD corrections, for the 8 TeV LHC, with \texttt{MadGraph5}~\cite{Alwall:2011uj},
\texttt{Pythia6}~\cite{Sjostrand:2006za} and \texttt{Delphes3}~\cite{deFavereau:2013fsa} after implementing the model in \texttt{FeynRules}~\cite{Alloul:2013bka}.
Jets are clustered with the shower-$k_T$ algorithm and jet matching is performed in the MLM scheme~\cite{Mangano:2006rw} at the scale $\frac{M_{D}}{4}$. 
We checked that differential jet rate distributions are smooth across the soft-hard jet threshold.

The processes of Eq.~(\ref{jjMET}) contribute to signatures with at least two hard jets and missing energy which mimic the production and 
decay of squarks and gluinos. Monojet signatures receive their main contributions from the process of Eq.~(\ref{jMET1}), with a subleading 
contribution from Eq.~(\ref{jMET2}) where the harder jet of the event is an initial state radiation QCD jet. Experimental searches for dark 
matter in monojet signatures are based on exclusive criteria to select events, discarding those events with two or 
more harder jets~\cite{Khachatryan:2014rra}. For this reason, processes like Eq.~(\ref{jjMET}), with at least two hard jets, contribute 
little to monojets.

Collider searches constrain the parameters related to the production cross section 
of the process discussed above. We have chosen to constrain the Yukawa coupling 
$y_D$ and the vector-like quark mass $M_D$, after fixing all the other parameters of 
the model.
We performed scans over a wide portion of the parameters space comprising $M_D$, $M_{H^{0}},M_{A^{0}},M_{H^{\pm}}$ and $y_{D}$. 
For each of these points we generated $10^4$ events for further analysis. The parameters scans were made as follows: \\

\begin{itemize}
\item First, with $M_{D}$ fixed, we varied $M_{H^{0}},M_{A^{0}},M_{H^{\pm}}$ and $y_{D}$ starting 
with $M_{H^{0}}=100\hspace{0.2cm}\hbox{GeV},M_{A^{0}}=M_{H^{\pm}}=105$ GeV until 
reaching almost the degeneracy of $D$ and the  scalars, always keeping the 
hierarchy $M_{D}>(M_{H^{0}}=M_{A^{0},H^{\pm}}-5$ GeV$)$, and varying the Yukawa 
couplings in the range $0.01\leq y_{D} \leq 1$;

\item Second, we varied $M_{D}$ in steps of 100 GeV starting with $M_{D}=300$ GeV up to $M_{D}=1.2$ TeV, proceeding with the first step for each $D$ mass. 
\end{itemize}

We used \texttt{SmodelS}~\cite{Kraml:2013mwa} to check for SUSY bounds and \texttt{CheckMate}~\cite{Drees:2013wra} for monojet bounds.
While the main input for \texttt{SModelS} is the full model definition given by the SLHA file containing masses, branching ratios and
cross-sections, \texttt{CheckMate} demands full simulated events to check for monojet bounds. 

We found that all scanned points passed the monojet constraints from \texttt{CheckMate}, but not from searches for hadronic decays of 
squarks and gluinos. \texttt{SmodelS}  decomposes the full model 
into simplified model spectrum topologies taking into account efficiency selection criteria in order
to make the correct comparison with its internal database. After that, it seeks for an experimental bound on the cross-section times
branching ratio, $\sigma(pp\to D_1D_2)\times BR(D_{1(2)}\to q+H^0)$ in our case, from 
a list of experimental publications and conference notes. Upper limits from those 
experimental studies on the cross sections, $\sigma_{95\%}$, at 95\% confidence 
level (CL), are then compared to the simulated $\sigma(pp\to D_1D_2)\times 
BR(D_{1(2)}\to q+H^0)$.
A model is considered excluded with CL above 95\%, for one or more analysis, 
whenever we have $\sigma(pp\to D_1D_2)\times BR(D_{1(2)}\to q+H^0) > \sigma_{95\%}$, or, in terms of the 
ratio $r\equiv \frac{\sigma(pp\to D_1D_2)\times BR(D_{1(2)}\to q+H^0)}{\sigma_{95\%}}$, if the output is $r\geqslant 1$. 

\begin{figure}[t]
\includegraphics[scale=0.42]{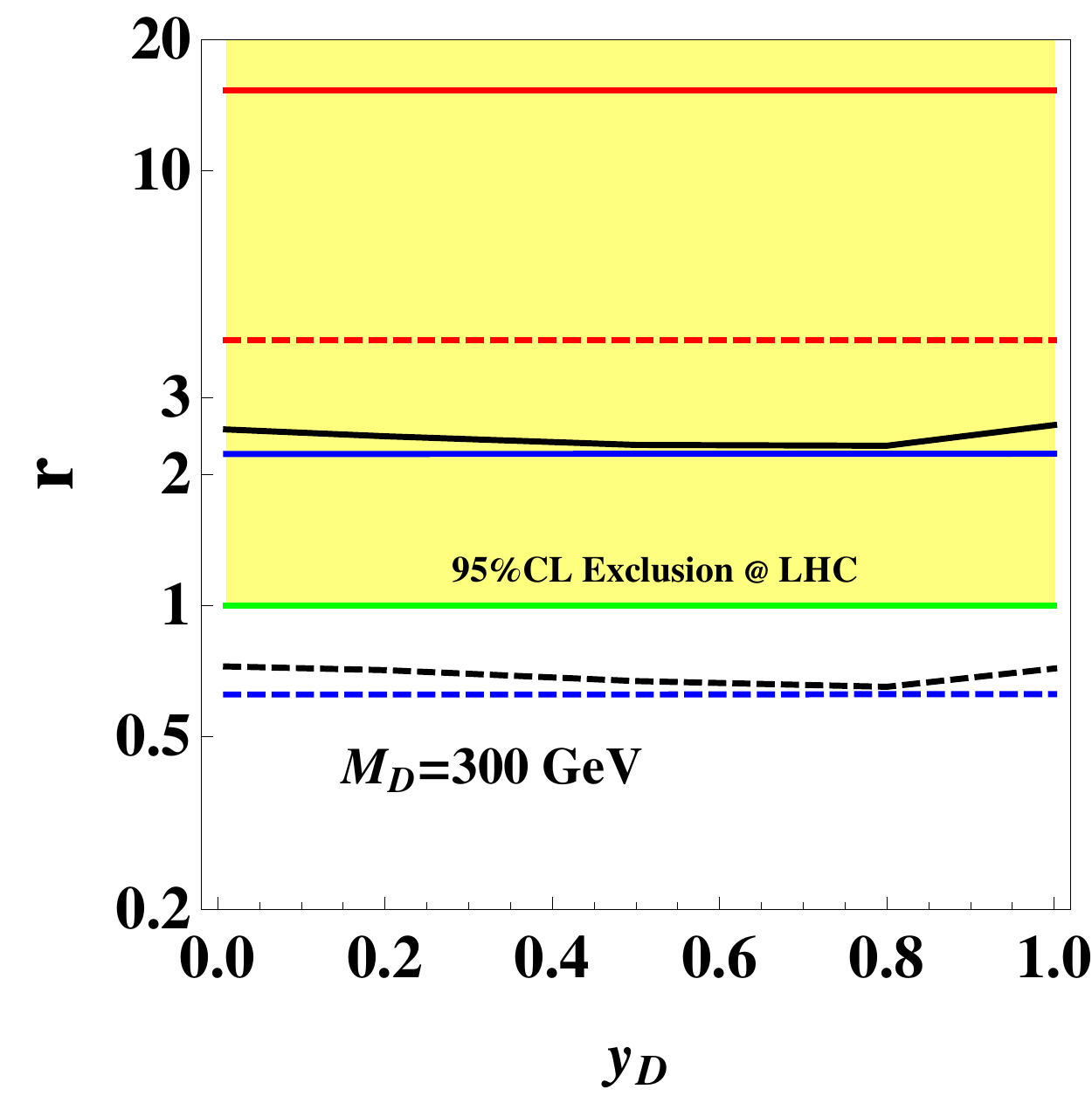}
\includegraphics[scale=0.42]{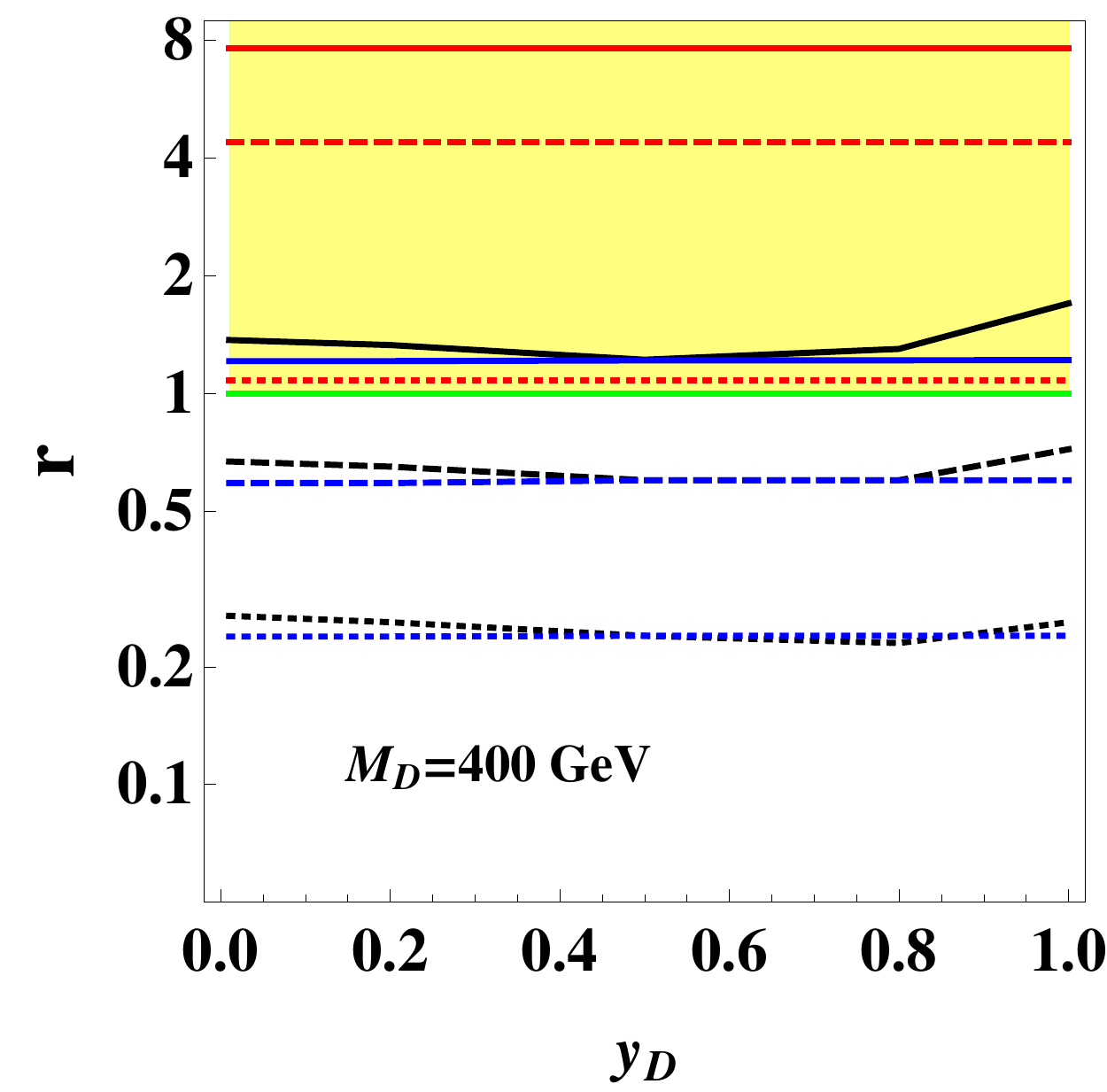} 
\includegraphics[scale=0.42]{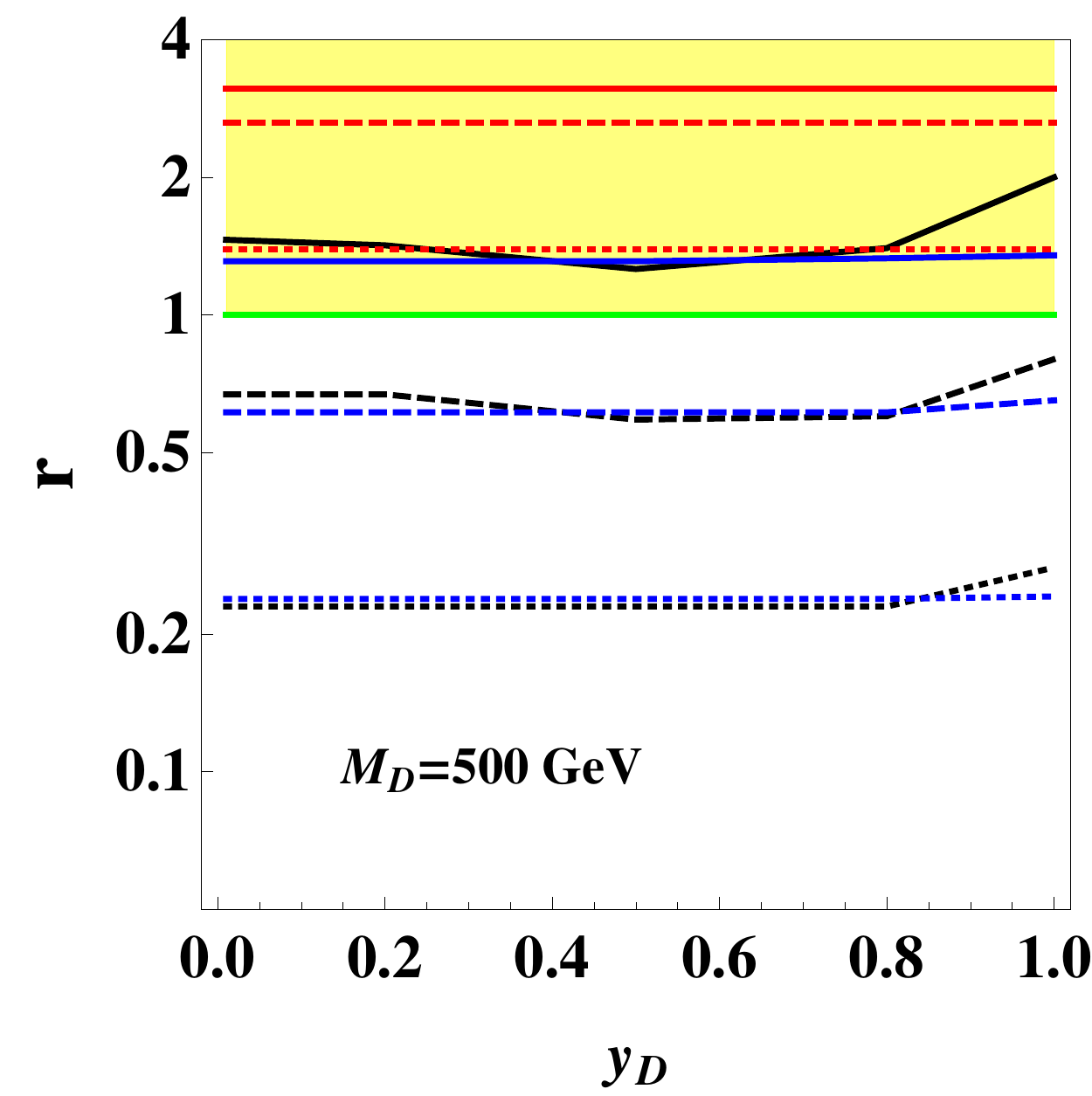}\\
\includegraphics[scale=0.42]{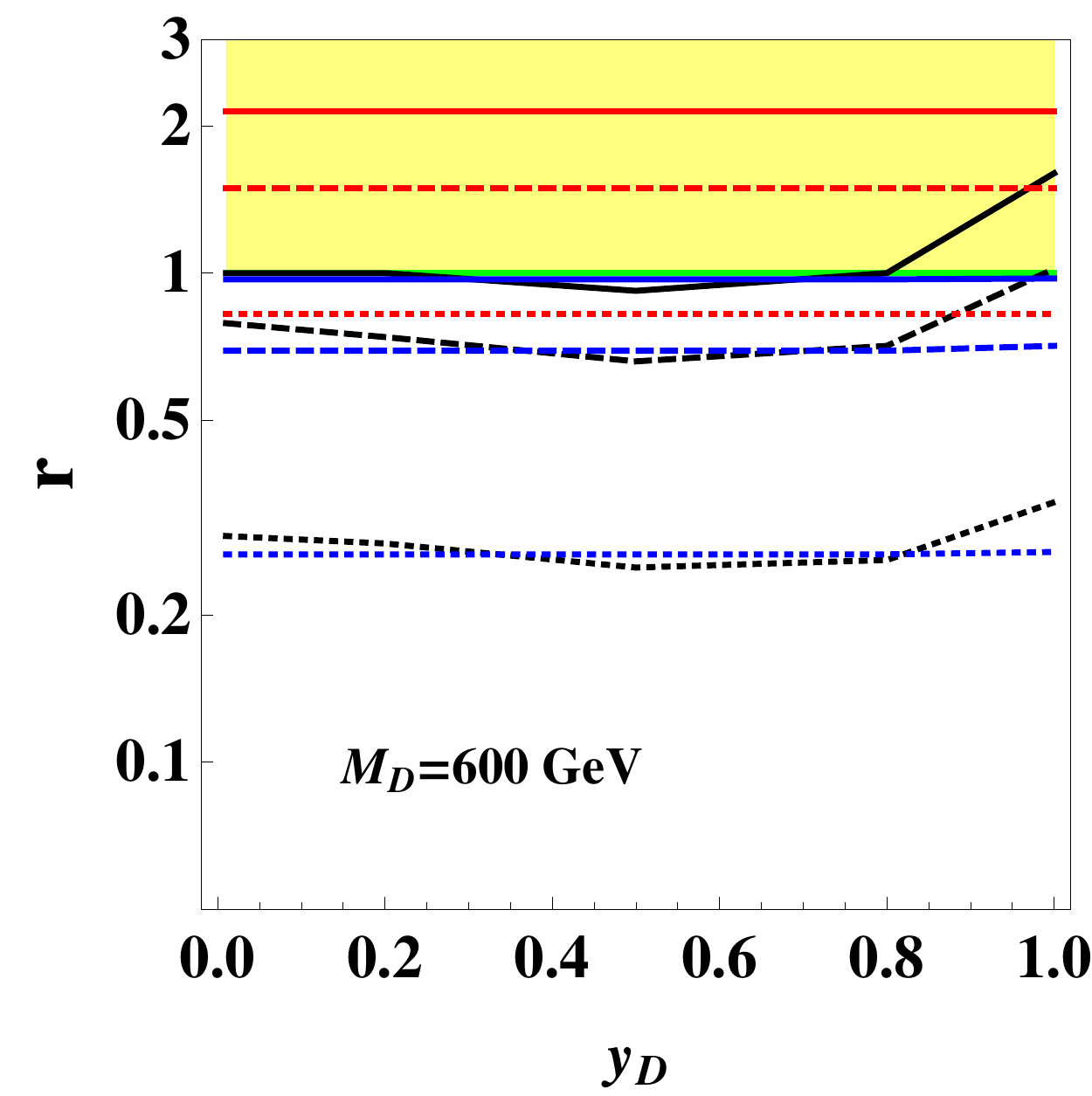}
\includegraphics[scale=0.42]{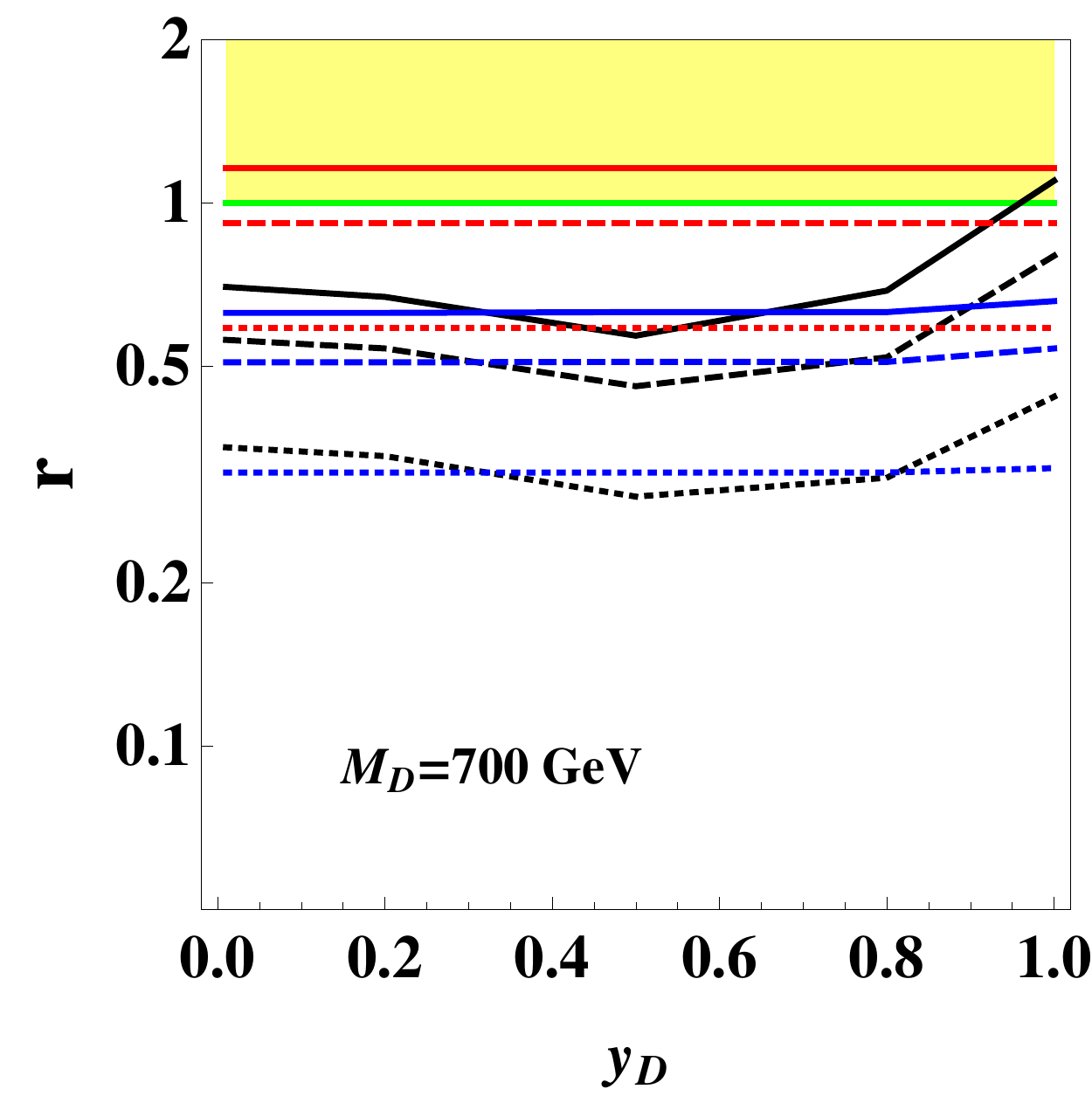}
\includegraphics[scale=0.42]{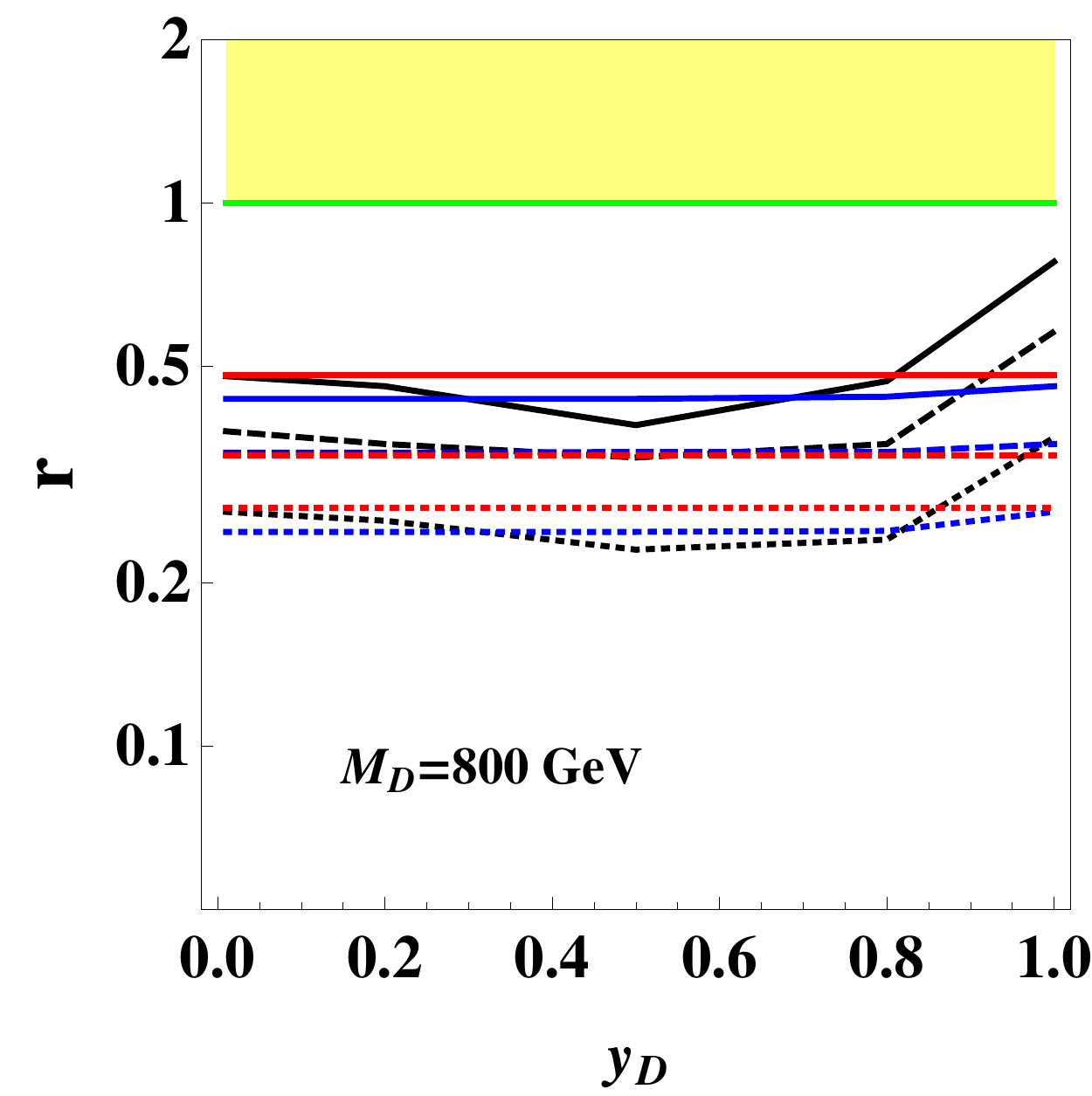}
\caption{Values of the ratio $r$, defined in the text, as a function of $y_{D}$ for 
several values of $M_{D}$. The shaded yellow area corresponding to $r\geq 1$ is 
excluded with 95\% at least. For each $M_{D}$, the solid, dashed and dotted lines 
correspond to $M_{H^{0}}=100$ GeV,  $200$ GeV, and  $300$ GeV, respectively. In all 
these scenarios, $M_{H^\pm}=M_{A^0}=M_{H^0}+5$ GeV, where black, blue and red lines 
correspond to $D$ coupling with the first, second and third family, respectively.} 
\label{fig:rscenarios}
\end{figure}

In Fig.~\ref{fig:rscenarios}, we show some possible scenarios 
corresponding to particular selections of the parameters of the model relevant for 
the DM analysis of the next section, where $D$ 
couples exclusively with either the first family (black lines), 
the second  family (blue lines) or the third  family (red lines).
For each scenario, the yellow shaded regions above $r=1$ can be considered excluded 
with 95\% CL, at least. 
For each $M_{D}$, the solid, dashed and dotted black lines correspond to the scenarios $(M_{H^{0}},M_{A^{0},H^{\pm}})=(100,105)\,\text{GeV}$, $(M_{H^{0}},M_{A^{0},H^{\pm}})=(200,205)\,\text{GeV}$ and $(M_{H^{0}},M_{A^{0},H^{\pm}})=(300,305)\,\text{GeV}$, respectively.

The first observation that we can draw from Fig.~\ref{fig:rscenarios} is that the most restrictive scenario occurs when $D$ interacts with the 
third family. In this case, $D$ has a typical SUSY signature matching with searches for direct production 
of bottom squark pairs which translates to harder constraints in our case. Second, 
the bounds for the second and third families are very weakly dependent on the Yukawa 
coupling, an effect that we have anticipated previously. On the hand, for Yukawa 
couplings between 0.2 and 0.8, approximately, first family scenarios have smaller 
$r$ ratios by virtue of the destructive interference between QCD and Yukawa 
contributions. Although the effect is not so pronounced, for $y_D\gtrsim 
0.8$ we see a clear trend towards the exclusion region as, in this regime, the 
production cross section increases.

We also see that, in general, as the mass of $D$ increases the production cross section drops fast as shown 
in Fig.~\ref{fig:xsex}, but the cut efficiency somewhat compensates for the signal decrease up to 700 GeV approximately as the jets becomes harder. For masses larger than $\sim 700$ GeV, the production cross section is too low and the model evades the collider constraints unless the Yukawa coupling is larger than 1.\\

In the next section we present the results of our analysis of the DM
candidate of the model taking into account all the collider constraints we obtained.

\section{Dark Matter Phenomenology }
\label{sec:DM}

Our work is based on a two component DM,   \textit{videlicet}, comprised of 
a WIMP ($H^0$) and an axion ($a$) (see 
\cite{Bae:2013hma,Bae:2014efa,Queiroz:2014ara,Bae:2015efa,Bae:2015rra} for other 
realizations of WIMP plus axion DM scenario). It is suitable to address 
the important aspect of the relic density of each component individually before 
discussing the WIMP+axion scenario. We start by reviewing how the WIMP abundance is 
obtained.

\subsection{WIMP Relic Density}

The abundance of the WIMP is obtained in the usual way, by solving the Boltzmann 
equation, with the help of 
\texttt{micrOMEGAS}~\cite{Belanger:2006is,Belanger:2008sj}. In this realization, the 
WIMP is in thermal equilibrium with SM particles, i.e., the annihilation and 
production interactions occur at similar rates in the early Universe. Although, as 
the Universe expands and the temperature drops below the DM mass, they can 
no longer be produced, and are simply able to pair-annihilate. 
Eventually, the expansion rate equals the rate for pair annihilation and then 
freeze-out is established.
Thus, the larger the annihilation cross 
section the fewer DM particles were left-over after the freeze-out. From 
then on, the abundance of left-over DM particle is kept basically constant.
This is the standard picture, where no coannihilations are present.
For the IDM, this is not the case and coannihilations play a dominant role in  
the WIMP abundance.

In the IDM the $H^0$ pair annihilation into SM particles is of the order of $6\times 
10^{-26}\,\mathrm{cm^3/s}$, for $500\,{\rm GeV} < M_H^0 < 3$~TeV 
\cite{Queiroz:2015utg}, which would naively produce an abundance below the correct 
value. Nevertheless, the other inert scalars $H^{\pm},A^0$ interact at similar rates 
with $H^0$ and SM particles, which makes them freeze-out at a similar time. Since 
they are not stable, after the freeze-out they decay into $H^0$ increasing its 
abundance to match the correct value. This mechanism was explained in detail 
recently in \cite{Queiroz:2015utg,Garcia-Cely:2015khw}. Thus, coannihilations are an 
important ingredient in the IDM in order to have a viable WIMP. The setup remains 
identical in the WIMP+axion framework that we will advocate, as long as the 
coannihilations involving the exotic quark $D$ are suppressed  (to be considered in 
section \ref{sec:coannihilations})
and axions have an insignificant relic density.

The IDM DM phenomenology can be wisely split into three mass regimes 
\cite{Barbieri:2006dq,LopezHonorez:2006gr,Cirelli:2005uq,Hambye:2009pw}\\

{\bf Low Mass: $M_{H^0} < M_W$}

In this mass range the model resembles the singlet scalar Higgs portal DM where the 
main annihilation modes are into light fermions, mainly $bb$ quarks, with 
annihilations controlled by the quartic coupling that mix the SM Higgs and $H^0$ 
\cite{Silveira:1985rk,McDonald:1993ex,Burgess:2000yq,O'Connell:2006wi,
Barger:2007im,Farina:2009ez,Kadastik:2011aa,Mizukoshi:2010ky,Alvares:2012qv,Djouadi:2012zc,Cline:2013gha,
Dasgupta:2014hha,Cogollo:2014jia,Alves:2014yha,Queiroz:2014pra,Feng:2014vea,Duerr:2015mva,
Duerr:2015aka,Beniwal:2015sdl,Han:2015hda,Dupuis:2016fda,Han:2016gyy}. There, $A^0$ 
and $H^{\pm}$ have to decouple in order to avoid direct and indirect DM 
WIMP searches. In summary, one needs to live at the Higgs resonance to be able to 
reproduce the right relic density while avoiding existing constraints 
\cite{Blinov:2015qva}.

{\bf Heavy Dark Matter: $M_{H^0} > 500$~GeV}

This mass region is viable and consistent with direct, indirect and collider searches. It can reproduce the right relic density thanks to coannihilations effects involving the inert scalars as we mentioned earlier, with a mass splitting $100\,{\rm KeV} < M_{H^0} -M_{H^+,A^0} < 15$~GeV \cite{Hambye:2009pw}. For recent studies in this mass region we refer to \cite{Queiroz:2015utg,Garcia-Cely:2015khw}. Interestingly, almost the entire parameter space of the model is expected to be probed by the Cherenkov Telescope Array \cite{Queiroz:2015utg,Garcia-Cely:2015khw}. 

{\bf Intermediate Mass: $M_W < M_{H^0} < 500$~GeV}

This mass region has been entirely excluded in the light of recent direct detection 
limits and relic density constraints \cite{LopezHonorez:2010tb}. Here, the 
annihilation rate into gauge bosons is very efficient leading to a dwindled relic 
density. It is in this precise mass region which the two component DM scenario we 
are advocating is most relevant. Since the WIMP share its duties with the axion, the 
constraints are relaxed and the total relic density of $\Omega_{total}=0.1$ can be 
achieved, motivating our work. We will explicitly show further how this is realized.

\subsection{Axion Relic Density}

As for the axion, the key question turns out to be, when is the Peccei-Quinn 
symmetry broken: before or after the inflation period? If it is broken before the 
end of inflation, the only process relevant for axion production is coherent 
oscillation due to the vacuum realignment and the axion relic density is given by 
\cite{Sikivie:2006ni, Bae:2008ue},

\begin{align}
  \label{eq:axionRelic}
 \Omega_a h^2 \sim 0.18 \theta^2\left(\frac{f_a}{10^{12}{\rm GeV}}\right)^{1.19},
\end{align}where $\theta$ is the initial axion misalignment angle. Note that, if 
$\theta$ is of order of unity, the axion can reproduce the total relic density, 
$\Omega_a h^2\sim \Omega h^2$, only for $f_a \sim 10^{12}$ GeV. We will set 
$\theta=1$ throughout.

In summary, the total relic density is given by $\Omega h^2 = \Omega_{H^0}h^2 + 
\Omega_ah^2 $, where $\Omega_{H^0}h^2$ is the relic density due to the WIMP, and 
$\Omega_ah^2$ the one corresponding to the axion, which depends on the cosmological 
model. That said, it is a good timing to discuss the two component DM abundance in 
more quantitative terms.

\subsection{Mixed WIMP-Axion Dark Matter in the IDM}
\label{sec:idm.limit}

\begin{figure}[t]
\centering
\includegraphics[scale=0.7]{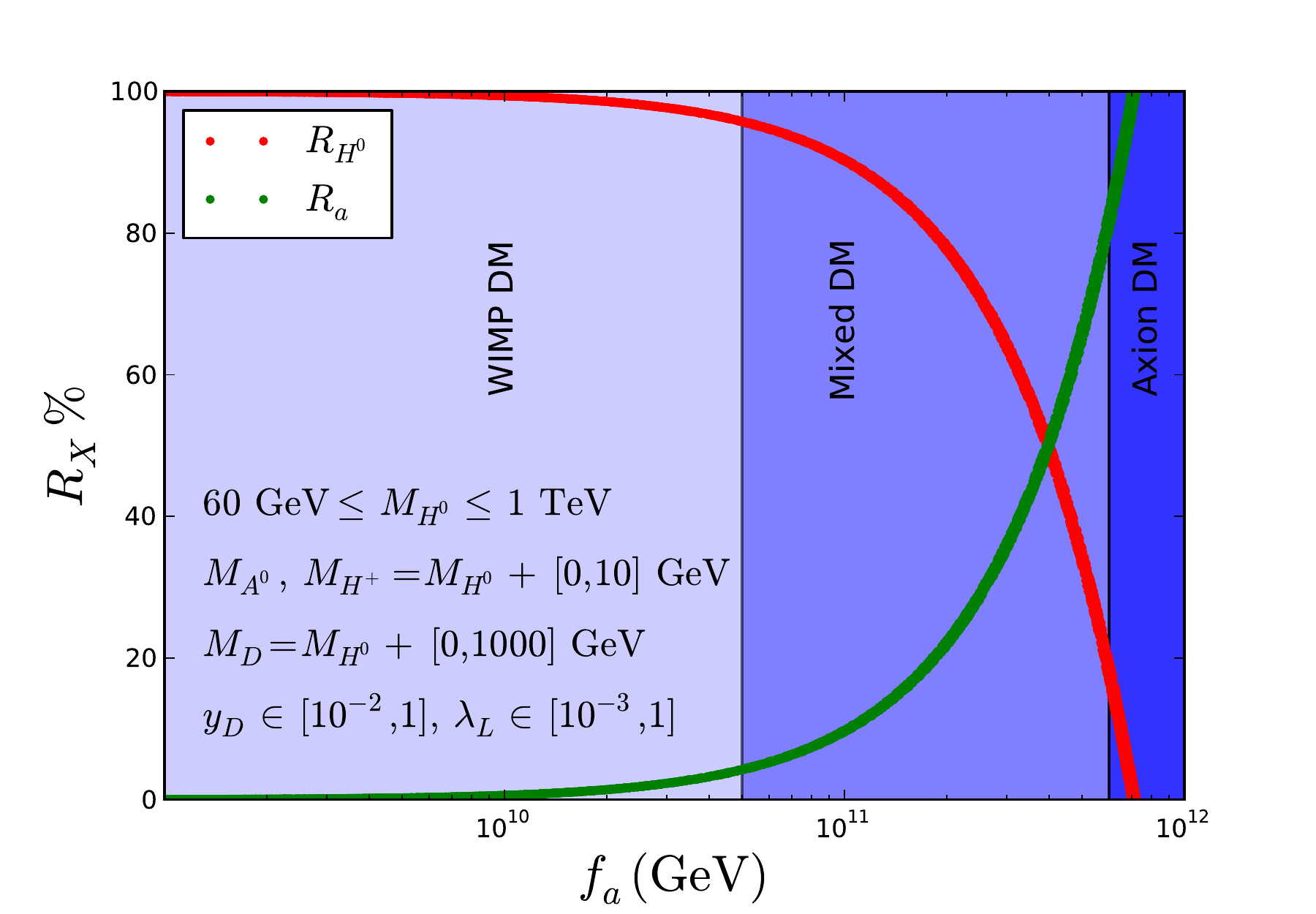}
\caption{Contributions to the total relic density $\Omega h^2 \sim 
0.1199\pm0.0027$~\cite{Komatsu:2010fb} as a function of the PQ scale $f_a$. The 
plot 
is the similar for our scenario in relation to the one  presented in 
Ref.~\cite{Dasgupta:2013cwa}. We have assumed $M_D > 300$ GeV, $\theta=1$, and the 
restrictions in Eq.\,\eqref{param:range}. The reason $H^0$ can meet the correct 
abundance is due to coannihilations involving the heavy vector-like quark.}
\label{fig:ratio}
\end{figure}

In order to take into account both axion and WIMP contributions to the total 
observed relic density\footnote{To calculate the WIMP contribution,
we have implemented the model in \texttt{FeynRules}~\cite{Alloul:2013bka} and used \texttt{microMEGAs}~\cite{Belanger:2013oya}.} 
we have scanned the free parameter space in the range shown in Table \ref{eq:scan}, 
always enforcing $M_{A^0}-M_{H^0}\gtrsim 100\,\unit{keV}$ to avoid the ruled out 
ineslatic DM regime \cite{TuckerSmith:2001hy,Arina:2009um}. The result of 
this scan is displayed in Fig.\;\ref{fig:ratio}.  There we show the relative WIMP 
and axion contributions to the total abundance as a function of $f_a$.   In 
Fig.\;\ref{fig:ratio} we have assumed that the exotic $D$ quark couples only to one 
family of SM quarks at a time through $y_D$, and concluded that the results are 
basically identical with a mild difference, within 3\%, for the third family, as 
one can see in Fig. \ref{fig:xsex}.

There important remarks are in order:

{\bf (i) } We can clearly see that for $f_a \lesssim  5 \times 10^{10}$~GeV, we 
enter the WIMP dominated regime.

{\bf (ii)} For  $5 \times 10^{10}\, {\rm GeV} \lesssim  f_a \lesssim 7 \times 
10^{11}$~GeV, we have a plausible two component DM setup being able to 
meet $\Omega_{total}h^2=0.12$.

{\bf (iii)} For $f_a > 7 \times 10^{11}$~GeV, we go into the axion dominated 
scenario. 

\begin{table}[t]
\eq{\nonumber
\begin{array}{|c|c|}
\hline 
\text{Parameter}  & \text{Scan range} \cr
\hline
M_{H^0} & 60 \text{ -- } 10^3\,\unit{GeV}
\cr
M_{A^0}-M_{H^0} & 0 \text{ -- }10\,\unit{GeV}
\cr
M_{H^\pm}-M_{H^0} & 0\text{ -- }10\,\unit{GeV}
\cr
\lambda_L & 10^{-3}\text{ -- }1
\cr
M_D -M_{H^0} & 0\text{ -- }10^3\, {\rm GeV}
\cr
y_D & 10^{-2}\text{ -- } 1
\cr 
f_a & 10^{9}\text{ -- }10^{15}\,\unit{GeV}
\cr
\hline
\end{array}
}
\caption{\label{eq:scan}
Parameter range used for the DM scan.
}
\end{table}

\begin{figure}[t]
\includegraphics[scale=0.7]{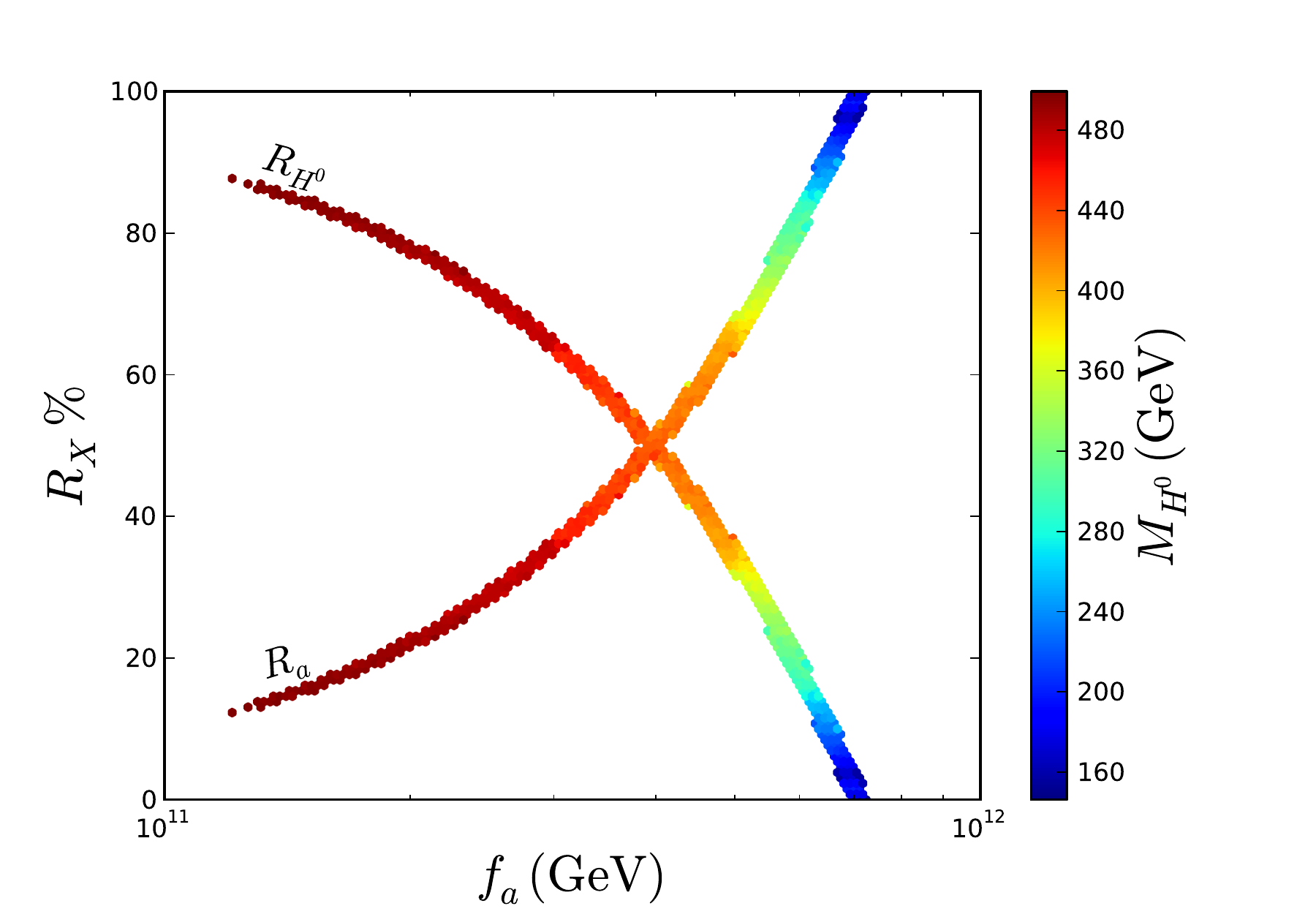}
\caption{Relative contribution of the inert scalar $H^0$ and axion to the total 
relic density, defined as $R_X$, as a function of $f_a$. The curve starting at $R_X 
> 80\%$ represents the inert scalar $H^0$ abundance, while the curve starting at 
$R_X <20\%$ reflects the axion.  We enforced the total relic density to be $\Omega 
h^2 \sim 0.1199\pm0.0027$~\cite{Komatsu:2010fb} throughout. We have assumed $M_D > 
300$ GeV, $\theta=1$, and the restrictions in Eq.\,\eqref{param:range}.}
\label{fig:ratio2}
\end{figure}

This plot visibly proves that one can successfully have a two component DM 
in the model. However, an important information in this two component DM 
scenario is the WIMP mass. That said,  we display in Fig.~\ref{fig:ratio2} the 
fractions $R_X$, with  $X=H^0,\,a$, of the total relic density as 
a function of the Peccei-Quinn scale $f_a$ explicitly showing the DM mass 
encoded in the curves. The fraction of relic abundance is defined as
\begin{align}
 \label{eq:ratio}
 R_X = \frac{\Omega_X h^2}{\Omega h^2}.
\end{align}
We have imposed $M_D> 300 $ GeV and the misalignment angle $\theta=1$.  In 
addition, we have also considered the constraints \eqref{param:range} discussed in 
the end of Sec.\,\ref{sec:model} and the restrictions showed in 
Fig.~\ref{fig:rscenarios}. The curve starting at $R_X > 80\%$ represents the inert 
scalar $H^0$ abundance, while the curve starting at $R_X <20\%$ reflects the 
axion's.  We enforced the total relic density to be $\Omega h^2 \sim 
0.1199\pm0.0027$~\cite{Komatsu:2010fb} throughout. 
We see clearly in Fig.\;\ref{fig:ratio2} that the WIMP dominated regime favors 
heavier masses ($M_{H^0} > 400$~GeV), 
whereas the axion dominated one prefers $M_{H^0} < 280$~GeV. The reason why the 
WIMP 
dominated region prefers heavier masses is just a consequence of the IDM nature of 
the WIMP, since it is well known that for $M_W <M_{H^0} < 500$~GeV the WIMP cannot 
produce $\Omega h^2 \sim 0.1199\pm0.0027$. As aforementioned, this is no longer 
problematic in the light of our two component DM where the axion abundance 
makes up for the deficit, depending on the value of $f_a$.

In Fig.\;\ref{fig:ratio} the WIMP can account for 100\% of the relic density as 
$f_a$ drops well below $10^{10}$~GeV, because there we entered in the mass region 
$M_{H^0}> 500$~GeV where the relic density constraint is satisfied. The heavy 
quarks also play a role in setting the WIMP abundance through coannihilation 
processes, when $M_D \sim M_{H^0}$, as we will investigate in detail further.  

\subsection{New Coannihilations with Vector-like Quarks}
\label{sec:coannihilations}

The DM phenomenology of the IDM from Peccei-Quinn symmetry differs from the IDM in 
two fundamental ways: (i) the presence of coannihilations involving the heavy 
vector-like quarks (D); (ii) the axion now contributes to the total relic density.

The new coannihilation processes involving the initial states $H^0 D$, 
$A^0 D$, $H^{\pm} D$ and $\overline{D} D$, will appear mediated by the Yukawa 
coupling 
$y_D$. Such coannihilations are exponentially suppressed by the mass splitting 
$\Delta M \equiv M_D - M_{H^0}$, and proportional to the Yukawa coupling $y_D$. If 
the mass difference is sufficiently large or the Yukawa coupling is dwindled, the 
$H_0$ phenomenology remains identical to the IDM. To quantify the impact of these 
new coannihilation processes on the WIMP relic density of the IDM from Peccei-Quinn 
symmetry, we have used the scan over the free parameters showed in Table 
\ref{eq:scan}. We have found that the coannihilation processes with the exotic 
quark 
$D$ 
are negligible when $M_D\gtrsim 1.2\, M_{H^0}$  and 
$y_D\lesssim 0.7$, so that we recover the DM phenomenology of the IDM in such a 
case, even though the coannihilation process $\overline{D} D \rightarrow gg$ has 
pure gauge contributions independently of the Yukawa $y_D$.

Generally speaking, coannihilation processes such as these only play a role if the 
mass splitting between the WIMP and the other odd particles is within 10-15\%, due 
to the Boltzmann suppression, which is the reason for negligible coannihilation 
processes when $M_D\gtrsim 1.2\, M_{H^0}$.

\begin{figure}[t]
\centering
\includegraphics[scale=0.44]{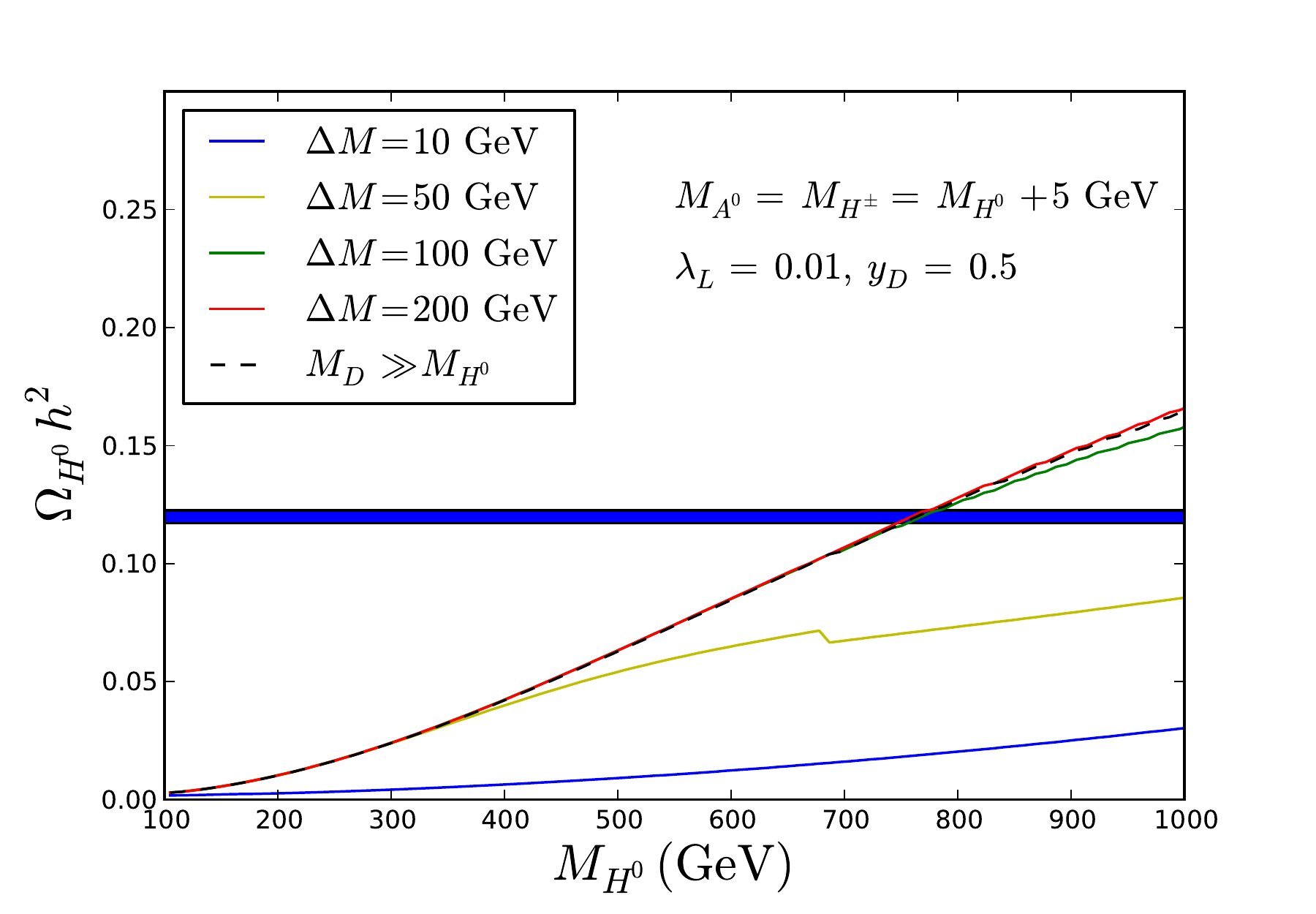}
\includegraphics[scale=0.44]{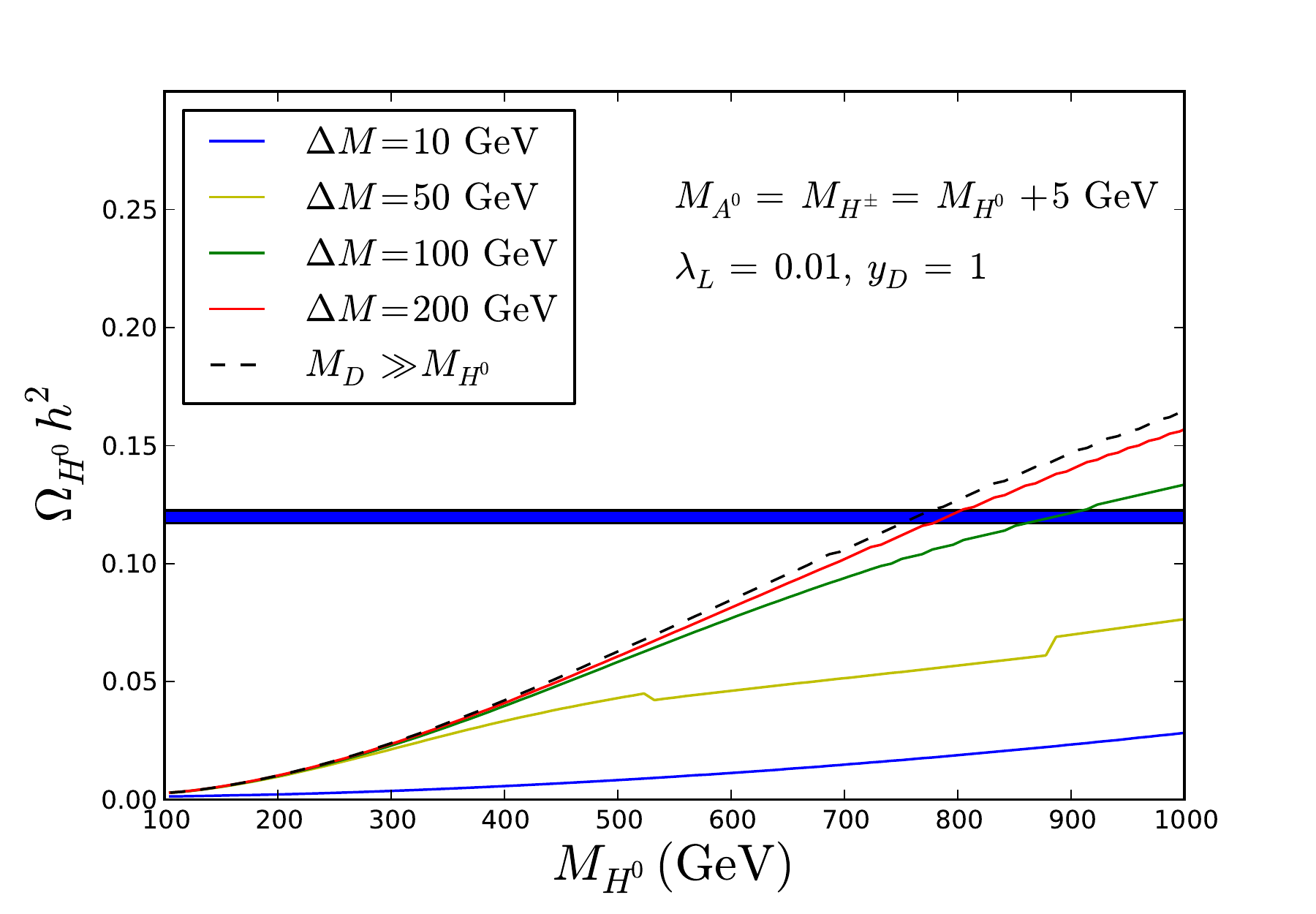}
\caption{WIMP relic density as a function of $M_{H^0}$ for different values of 
$\Delta M \equiv M_D - M_{H^0}$ 
and for $y_D=0.5$ (left panel) and $y_D=1$ (right panel).
The horizontal blue line correspond to the actual experimental bound $\Omega h^2 
\sim 0.1199\pm0.0027$ \cite{Komatsu:2010fb}.
The decoupling limit, $M_D\gg M_{H^0}$, coincides with the inert doublet model.}
\label{fig:coa}
\end{figure}

We display in Fig.\;\ref{fig:coa} the WIMP relic density as a function of $M_{H^0}$ 
for the mass differences $\Delta M = 10$ GeV (blue line), 50 GeV (yellow line), 100 
GeV (green line), 200 GeV (red line)
and for two values of the Yukawa coupling $y_D = 0.5$ (left panel) and $y_D = 1$ 
(right panel). 
The dashed line correspond to the decoupled limit, $M_D\gg M_{H^0}$, where the 
coannihilations are negligible and 
the IDM phenomenology is recovered. 
The horizontal blue band correspond to the current bound $\Omega h^2 \sim 
0.1199\pm0.0027$ \cite{Komatsu:2010fb}.

Note that the coannihilations with the exotic quark decrease the WIMP population
and increase the allowed DM mass compatible with the data.
That is because the inclination of the relic density curve of $H^0$ depends 
on how efficient  vector-quark coannihilations are.
Thus, once we reach the overabundant regime, we 
can simply {\it turn on} such coannihilation by increasing $y_D$ and making the 
mass difference smaller, and bring down the abundance to the correct vale. In 
other words, we simply change the inclination of the abundance curves as can be 
explicitly seen in Fig. \ref{fig:coa}. 

In particular, for $y_D=1$, right panel of Fig. \ref{fig:coa}, we can see a 
significant difference between the case in which $\Delta M = 200$ GeV (red line), 
where the WIMP reproduce the total relic density for $M_{H^0}\approx 800$ GeV, and 
the case in which $\Delta M = 100$ GeV (green line), where the WIMP reproduce the 
total relic density for a larger mass of $M_{H^0}\approx 900$ GeV. It is only for a 
splitting $\Delta M > 200$ GeV that our model recovers the IDM phenomenology, where 
the vector-like quark coannihilations are turned off. For $y_D=0.5$ this mass 
difference is $\Delta M > 100$ GeV. Notice that for $y_D = 1$, the coannihilation 
cross sections are larger and hence a mass splitting must be mildly larger compared 
to the case with $y_D = 0.5$ in order to suppress the coannihilations, where 
$\Delta M > 100$ GeV suffices.

In the collider section we observed that $y_D > 0.8$ might be problematic due to 
monojet and dijet plus missing energy constraints, therefore $y_D=0.5$ is a 
feasible benchmark model, where both relic density and collider constraints are 
satisfied as well as the direct and indirect DM detection probes addressed 
in the following.

\subsection{Direct Detection}

WIMPs might also scatter off of nuclei and deposit an energy which can be measured 
by underground detectors such as LUX \cite{Akerib:2015rjg}, CDMS 
\cite{Agnese:2014aze} and PICO \cite{Amole:2015lsj} among others 
\cite{Agnes:2014bvk,Xiao:2014xyn,Archambault:2012pm,
Aartsen:2016exj,Choi:2015ara,Mijakowski:2016cph}, 
all of them using different target nuclei and readout techniques. 
By discriminating nuclear recoil from electron recoils, the experiments have been 
able to place stringent limits in the scattering cross section vs WIMP mass, 
capable of depositing an energy above few keV. In the IDM model, the direct 
detection limits from LUX, which is currently the world leading experiment, can be 
easily evaded by requiring the mass splitting 
between $A^0$ and $H^0$ to be above $100$~KeV, and the coupling $\lambda_L$ to be 
suppressed with no prejudice to our reasoning.
In particular, the values $|\lambda_L|\lesssim 0.01$ are well below 
the current sensitivity of LUX \cite{Blinov:2015qva} and also the projected 
sensitivity of XENON1T \cite{Aprile:2012zx,Aprile:2015uzo}.

In our model augmenting the IDM, we need to consider the presence of the exotic 
quark $D$ which can mediate the WIMP interaction with the nucleus by 
s-channel and t-channel scattering with quarks/gluons, as 
shown in Fig.\;\ref{fig:production}, diagrams (g) and (h).%
\footnote{A study at one loop was realized in 
\cite{Giacchino:2015hvk} 
for the singlet scalar model augmented with a exotic quark and neglecting the Higgs 
portal.} Such interactions are governed by the Yukawa coupling $y_D$ and the 
exotic quark mass $M_D$.  When $M_D \sim M_{H^0}$, which is of interest to us since 
coannihilations with the $D$-quark become important, there is an enhancement in the 
cross section as a result of the inelastic regime. Taking $y_D \lesssim 0.5$, the 
model is consistent with the LUX bound on the spin-independent scattering cross 
section. Thus, from the left panel of Fig.\;\ref{fig:coa}, we can see that the 
model 
can simultaneously  yield the right abundance and accommodate the LUX limit. For 
$M_D \sim 1.2 \,M_{H^0}$ ($\Delta M\sim 0.2 M_{H^0}$), when coannihilations are 
turned off, we find that for $y_D \lesssim0.7$, the model is below LUX and future 
XENON1T \cite{Aprile:2012zx} bounds. In summary, our benchmark model with $y_D=0.5$ 
is perfectly consistent with current and projected limits from direct detection.

Thus we conclude that the right panel in Fig.\;\ref{fig:coa}, where $y_D=1$, is 
excluded in the light of direct detection experiments. This conclusion shows 
the high degree of DM complementarity in our model. However, this holds 
true as long as $H^0$ accounts for the total DM abundance, which is not 
necessarily true in our model, specially when $M_W <M_{H^0} < 500$~GeV.
Since the direct detection limits are linearly proportional to the WIMP local 
density, the bounds are alleviated and the model can be made compatible with direct 
detection in the regime where the axion makes up a large fraction of the 
abundance, i.e., for $f_a \gtrsim 7 \times 10^{11}$~GeV.
We handpicked these two values for $y_D$ to show precisely when direct 
detection constraints become relevant and how our two component DM scenario 
plays an important role in satisfying both relic density and direct detection 
searches for WIMPs.

\subsection{Indirect Detection}

WIMPs may self-annihilate producing a sizable amount of gamma-rays and cosmic-rays 
over the astrophysical background (see 
\cite{Profumo:2013yn,Queiroz:2016awc,Bertone:2016nfn} for recent reviews). Searches 
for WIMP annihilations have been performed in several target regions such as the 
Galactic Center, Dwarf Galaxies, Cluster of Galaxies etc. 
In our model the mass of interest is hardly touched by current Fermi-LAT and 
H.E.S.S. limits \cite{Abramowski:2011hc},
since the need for the axion to complement the WIMP under-abundance relaxes the 
indirect detection limits which depend on the local DM density squared.

Even assuming that $H^0$ makes up the entire DM of the Universe, for 
$500\;{\rm GeV} < M_{H^0} < 3$~TeV, Fermi-LAT limits are rather weak, with H.E.S.S. 
ruling just a tiny fraction of the parameter space \cite{Queiroz:2015utg}, unless 
boost factors are advocated \cite{Garcia-Cely:2015khw}. It is worth mentioned that 
the Cherenkov Telescope Array might improve existing limits in more than one order 
of magnitude, and depending on the level of systematics uncertainties achieved 
\cite{Lefranc:2015pza,Silverwood:2014yza}, the entire model below $3$~TeV might be 
excluded \cite{Queiroz:2015utg}. We emphasize though, that in our two component 
DM scenario such conclusions are strongly relaxed. In other words, our 
results are consistent with exclusion limits from indirect DM detection 
searches.

\section{Conclusions}
\label{sec:concluisions}

Since WIMPs and axions are arguably the most compelling DM candidates in 
the literature, we investigate the possibility of two component DM in a 
well motivated model, namely the Inert Doublet Model. We present a model that 
contains, beyond the SM fields, a scalar  inert doublet, a scalar singlet hosting an 
axion, and a new vector-like quark $D$. These fields allow an implementation of the 
Peccei-Quinn $U(1)_{PQ}$ symmetry that solves the strong CP problem and 
gives rise to an invisible axion. The inert doublet originates a candidate for dark 
matter, stabilized by a natural $\ZZ_2^D$ symmetry remnant from the breakdown of  
$U(1)_{PQ}$ symmetry following Ref.\,\cite{Dasgupta:2013cwa}.  The new quark 
provides a new portal to connect the SM to the dark sector, which is comprised of 
particles that are odd under $\ZZ_2^D$ 
transformations plus the axion. 

Along with the WIMP, the new quark gives rise to signals involving jets plus missing 
energy, and also monojets at the LHC. In order to investigate possible restrictions 
on the parameter space of the model, we have studied all these potential signals at 
the LHC considering that the $D$ quark couples to the WIMP and with just one of the 
SM families. We found that the most restrictive scenario occurs when $D$ quark 
couples to the third family bottom quark. For example, such a scenario is excluded 
at 95\% C.L for masses of the scalars $H^{0}$, $A^{0}$, and $H^{\pm}$ being 
$(M_{H^{0}},M_{A^{0},H^{\pm}})\leq(200,205)$ GeV, if $M_D\leq 600$ GeV and $y_D\leq 
1$. In the case where the $D$ quark couples with the first or the second family, the 
restrictions are milder, and masses $(M_{H^{0}},M_{A^{0},H^{\pm}})\geq(200,205)$ GeV 
are allowed for $M_D\geq 400$ GeV for all Yukawa couplings up to at least unity. 

In our model, DM is composed by two components, the lightest inert scalar 
($H^0$) and the axion. Within this scenario we performed an investigation on how the 
fractions of the DM relic abundance corresponding to the WIMP and to the 
axion change depending on the scale $f_a$ of the breakdown of the $U(1)_{PQ}$ 
symmetry, the mass of the WIMP,  the masses of the other particles odd by the 
$\ZZ_2^D$ symmetry. For example, for values $f_a\leq 10^{10}$ GeV the  WIMP would 
constitute essentially all the DM, with the axion being an irrelevant 
fraction of it. As $f_a$ increases the axion relic density raises, reaching a 
value equal to the  WIMP relic density for $f_a\simeq 4\times 10^{11}$ GeV. 

In contrast with the inert Higgs doublet model, we found that in our model it is 
possible to have the WIMP from the inert doublet with mass in the interval
$100\,\mathrm{GeV}\lesssim M_{H^0}\lesssim 500\,{\rm GeV} $, and comprising only 
a fraction of the total DM relic abundance. This region is 
phenomenologically important for direct detection experiments and LHC searches of
exotic quarks and DM. 
In particular, we have shown that the IDM phenomenology remains unchanged when the 
coannihilations effects with the exotic quark are negligible and this happens for 
$M_D\gtrsim1.2M_{H^0}$. We conclude that one can have a plausible two component 
DM satisfying the relic density as well as collider, direct and indirect 
DM detection constraints.

\section{Prospects}

The assumption that the DM is composed by two or more type of particles 
impacts on the experiments searching for  WIMPS and axions. For example, if the 
axion relic density  constitute an irrelevant fraction of the DM the axion 
could not be direct detected in haloscopes experiments~\cite{Asztalos:2011bm}, but 
it could still be accessible in the projected experiment 
IAXO~\cite{Armengaud:2014gea}, which arises as a promising laboratory to test 
the model we proposed.
On the WIMP side, direct future experiments with large 
exposure such as XENON1T 
\cite{Aprile:2015uzo} and LZ \cite{Malling:2011va} are quite desired. Future 
collider constraints stemming from a possible 100 TeV collider or linear collider 
might also constrain the model even further 
\cite{Barger:2013ofa,Wang:2014lta,Hajer:2015gka,Arkani-Hamed:2015vfh,
Golling:2016gvc}.

\section*{Acknowledgments}

A.\ Alves, A.G.\ Dias and C.C.\ Nishi acknowledge financial support from the 
Brazilian CNPq, processes 307098/2014-1, 303094/2013-3 and 311792/2012-0, 
respectively, and FAPESP, process 2013/22079-8 (A.A., A.G.D., C.C.N.). 
D.~Camargo thanks CAPES for financial support. R.~Longas is supported by 
COLCIENCIAS and acknowledges the hospitality of Universidade  Federal do ABC in the 
early stage of this work. F.\ S.\ Queiroz is grateful to the Mainz Institute for 
Theoretical Physics (MITP) for its hospitality and its partial support during the 
completion of this work.

\appendix

\section{Simple UV completions}

\subsection{$U(1)_{\pq}$ breaking in the Higgs potential}
\label{ap:uv}

It is natural to expect that the $U(1)_{\rm PQ}$ breaking at the high 
scale (larger than $10^9\,\unit{GeV}$) induces at lower energies the operator
in \eqref{lambda5}.
We present here a simple model where that happens.

To complete the model, we add another SM singlet scalar $\varphi$ with PQ 
charge unity but inert (no vev).
The relevant terms in the Lagrangian above the PQ scale will be
\eqali{
\label{lag:UV.1}
\lag &\supset \bar{q}_{L}H_D D_R+
S^*\overline{D}_L D_R+\varphi^*\overline{D}_Ld_{R}
\cr&\quad
+\ S^*\varphi^2+(H^\dag H_D)\varphi+(H^\dag H_D)\varphi^* S+h.c.
}
We omit the coefficients for simplicity and, for definiteness, we take the exotic 
quark to be of charge $-1/3$, denoting it by $D$. The case of charge $2/3$ is 
analogous.
The PQ charges are given in table \ref{tab:UV.1}.
\begin{table}[h!]
\begin{center}
\begin{tabular}{|c|c|c|c|c|c|}
\hline\rule[0cm]{0cm}{.9em}
  & $D_L$ & $ D_R$ & $H_D$ & $S$ & $\varphi$\\
\hline
\hline
 $U(1)_{PQ}$ & $-1$  & $1$ & $-1$  & $2$ & $1$ \\ \hline
\end{tabular}
\end{center}
\caption{Fields with nonzero PQ charges.}
\label{tab:UV.1}
\end{table}

After $S$ acquires a vev $\aver{S}$, the breaking $U(1)_{\pq}\to \ZZ_2^D$ is 
induced and we get effectively
\eqali{
\label{lag:UV.2}
\lag &\supset y^i_{D}\bar{q}_{iL}H_D D_R+
M_D\overline{D}_L D_R+\kappa_{j}^*\varphi^*\overline{D}_Ld_{jR}
\cr&\quad
+\ \mu_\varphi^{2}\varphi^{2}+\mu_H(H^\dag H_D)\varphi+
\mu_{H}'(H^\dag H_D)\varphi^*+\lambda_\varphi \varphi^4+ h.c.
}
We assume $\sqrt{|\mu^2_\varphi|}$ and the mass accompanying $|\varphi|^2$ 
to be much smaller than the PQ scale but much larger than the electroweak scale.
The $\varphi^2$ term splits the complex scalar into two real scalars 
$\varphi_1,\varphi_2$ of different masses $M_1,M_2$.
Thus the terms with $\mu_H,\mu_H'$ of \eqref{lag:UV.2}, which can be recast in the 
form
\eq{
(H^\dag H_D)(\mu_1\varphi_1+i\mu_2\varphi_2),
} 
leads to the desired operator 
\eqref{lambda5} with coefficient
\eq{
\lambda_5=\Big(\frac{\mu_1^2}{M_1^2}-\frac{\mu_2^2}{M_2^2}\Big)\,.
}

This model at the PQ scale is identical to the model I presented in 
Ref.\,\cite{Dasgupta:2013cwa} which realizes $U(1)_{\pq}\to \ZZ_2^D$ in a KSVZ type axion 
model and, additionally, also generates neutrino masses radiatively.
At the TeV scale, however, our focus is on a different physical spectrum 
where the DM candidates are the axion and the lightest neutral member of the inert 
doublet while the interaction of the heavy quark with the SM occurs also 
through the inert doublet.
We should also emphasize that a different realization may lead to the same physical 
spectrum at the TeV scale -- the SM augmented by an inert doublet, an axion and a 
exotic quark -- but to a different particle content at the PQ scale.

\subsection{Lighter exotic quark mass}
\label{ap:uv-qft}

For the model \eqref{lag:UV.2}, it is expected that the exotic quark mass $M_D$ be 
at the order of the PQ breaking scale or at most few orders of magnitudes lower. 
To get $M_D$ at the TeV scale one has to tune the Yukawa coupling to at least 6 
orders of magnitude. Here we show a variant where the exotic $D$ quark have mass decoupled from the PQ 
scale and thus can be lighter.

The variant includes another exotic quark, which we keep denoting as $D$, while the 
original exotic quark is renamed as $D'$.
Thus the new exotic quark $D$ has the same quantum numbers as $D'$ except that it 
is vector-like with respect to PQ symmetry: $\pq(D_L)=\pq(D_R)=1$.
Now $D$ is still the quark that couples to the SM quarks but the QCD anomaly is 
generated by $D'$.

The relevant Lagrangian is modified to
\eq{
    \label{lag:UV.3}
\lag\supset \bar{q}_L H_DD_R+S^*\overline{D'}_LD_R'+S^*\overline{D'}_LD_R
+\overline{D}_LD_R'+\overline{D}_LD_R+h.c.
}
After PQ breaking we get
\eq{
\lag\supset M_{D'D'}\overline{D'}_LD_R'+M_{D'D}\overline{D'}_LD_R
+\tilde{M}_{DD'}\overline{D}_LD_R'+\tilde{M}_{DD}\overline{D}_LD_R+h.c.,
}
where the coefficients are now explicitly written and the masses denoted by tilde 
are bare and in principle can be much smaller than the PQ scale.

We can write 
\eq{
\mathbb{M}_D=
\mtrx{\tilde{M}_{DD} & \tilde{M}_{DD'}\cr
    M_{D'D} & M_{D'D'}\cr
    }\,.
}
It is easy to see for the case of $\tilde{M}_{AB}\ll M_{AB}$, $A,B=D,D'$, that 
$U_L$ diagonalizing $\mathbb{M}_D\mathbb{M}_D^\dag$ has a small mixing angle while 
$U_R$ diagonalizing $\mathbb{M}_D\mathbb{M}_D^\dag$ has a large mixing angle.
After, integrating out the heaviest state, we end up with a lighter exotic quark 
with mass $M_D\sim O(\tilde{M}_{AB})$ with appreciable coupling to the SM quarks 
through the first term of \eqref{lag:UV.3}.

\bibliography{darkmatter}

\end{document}